\documentclass[%
 aps,
 prapplied,
 amsmath,amssymb,
 reprint,%
floatfix,
]{revtex4-2}

\usepackage{graphicx}
\usepackage{dcolumn}
\usepackage{bm}

\usepackage{xr-hyper}

\usepackage[utf8]{inputenc}
\usepackage[T1]{fontenc}
\usepackage{mathptmx}
\usepackage{etoolbox}
\usepackage{amsmath}
\usepackage{siunitx}

\DeclareMathOperator{\im}{Im}
\newcommand{\dd}{\mathrm{d}}

\usepackage[
  colorlinks=true,
  citecolor=blue,     
  linkcolor=blue,     
  urlcolor=blue,      
  breaklinks=true
]{hyperref}

\usepackage[acronym]{glossaries}
\glsdisablehyper 
\newacronym{dpph}{DPPH}{2,2-diphenyl-1-picrylhydrazyl}
\newacronym{epr}{EPR}{Electron Paramagnetic Resonance}
\newacronym{cad}{CAD}{computer aided design}
\newacronym{hemt}{HEMT}{high electron mobility transistor}
\newacronym{vna}{VNA}{vector network analyzer}
\newacronym{do}{DO}{digital output}
\newacronym{di}{DI}{digital input}
\newacronym{ao}{AO}{analog output}
\newacronym{ai}{AI}{analog input}
\newacronym{mxc}{MXC}{mixing chamber}
\newacronym{still}{Still}{distillation}
\newacronym{cp}{CP}{cold plate}
\newacronym{mw}{MW}{microwave}
\newacronym{lo}{LO}{local oscillator}
\newacronym{rf}{RF}{radio frequency}
\newacronym{if}{IF}{intermediate frequency}
\newacronym{cw}{CW}{continuous wave}
\newacronym{iq}{IQ}{in-phase and quadrature}
\newacronym{ssb}{SSB}{single sideband}
\newacronym{fid}{FID}{free induction decay}

\begin{document}


\title{Controlling the Inhomogeneous Broadening and Impedance Matching of a Spin Ensemble}
\author{Mathieu Couillard}
\affiliation{Experimental Quantum Information Physics Unit, Okinawa Institute of Science and Technology, Okinawa, Japan}

\author{Debdip Guchait}
\altaffiliation[Present address: ]{Faculty of Physics, Technion - Israel Institute of Technology, Haifa, Israel}
\affiliation{Experimental Quantum Information Physics Unit, Okinawa Institute of Science and Technology, Okinawa, Japan}

\author{Hiroki Takahashi}%
\affiliation{Experimental Quantum Information Physics Unit, Okinawa Institute of Science and Technology, Okinawa, Japan}

\author{Yuimaru Kubo}%
\affiliation{Science and Technology Group, Okinawa Institute of Science and Technology, Okinawa, Japan}
\email{yuimaru.kubo@oist.jp}

\date{\today}


\begin{abstract}
We control the spectral distribution of a spin ensemble by applying a magnetic field gradient using an anti-Helmholtz coil inside a dilution refrigerator, and demonstrate impedance matching between the ensemble and a transmission line, achieving $-50$~dB absorption of incident radiation.
This represents a first step toward a spin-ensemble-based quantum memory for itinerant microwave photons.
We further model the spectral distribution under the applied gradient to predict the spin--resonator response, and use the device to systematically tune the weak-to-strong coupling transition in both continuous-wave and time-domain pulsed measurements.
\end{abstract}

\maketitle

\section{Introduction}
Ensembles of quantum emitters play a central role in a wide range of quantum technologies and have been explored for applications such as quantum memories
\cite{lvovsky2009optical, gorshkov2007photon, afzelius2009multimode, morton2008solid, wu2010storage, kubo2011hybrid, zhu2011coherent, julsgaard2013quantum, afzelius2013proposal, grezes2014multimode, ranjan2020multimode, greggio2025optimal}, transducers\cite{lambert2020coherent, hafezi2012atomic, o2014interfacing, williamson2014magneto, fernandezgonzalvo2015coherent, hisatomi2016bidirectional, fernandezgonzalvo2019cavity, barnett2020theory, bartholomew2020chip, li2023fiber, rochman2023microwave, xie2025scalable}, near-quantum-limited amplifiers\cite{sherman2022diamond, day2024room, ohta2025near}, electromagnetic field sensors\cite{degen2017quantum, barry2020sensitivity, michl2019robust, fan2015atom, hong2013nanoscale}, simulators\cite{zhang2017observation,schafer2020tools}, atomic clocks\cite{essen1955atomic, aeppli2024clock}, and gyroscopes\cite{jarmola2021demonstration, soshenko2021nuclear}. 
They also provide a platform for the study of collective quantum phenomena such as superradiance\cite{dicke1954coherence, rose2017coherent, angerer2018superradiant}, subradiance\cite{guerin2016subradiance}, and self-stimulated echo trains\cite{weichselbaumer2020echo, debnath2020self}.

A particularly promising solid-state platform consists of spin ensembles coupled to microwave resonators, providing an interface that can transfer quantum information from superconducting and semiconducting quantum circuits via microwave photons.
These systems can exploit the long coherence times of spin ensembles\cite{bar2013solid, wolfowicz2013atomic, tyryshkin2012electron, le2021twenty} through strong collective coupling to microwave photons\cite{kubo2010strong, schuster2010high, amsuss2011cavity, abe2011electron, sandner2012strong, ranjan2013probing, bushev2011ultralow}, making them attractive for high-fidelity quantum random-access memory\cite{julsgaard2013quantum, wu2010storage, grezes2014multimode, o2022random, ranjan2022spin, ranjan2020multimode} and processing.

A key factor governing the dynamics of such systems is the spectral distribution of the ensemble. Its characteristic width is the inhomogeneous linewidth $\Gamma$, which sets how rapidly the ensemble dephases when prepared in a superposition.
Typically, the inhomogeneous broadening of a spin ensemble is treated as a fixed property, determined by local field variations due to nearby paramagnetic centers\cite{bauch2018ultralong, bauch2020decoherence}, crystal strain\cite{doherty2011theory, jamonneau2016competition, ranjan2020electron,
ranjan2021spatially, pla2018strain}, and temperature variations\cite{acosta2010temperature, toyli2013fluorescence} across the sample.

The dynamics of the coupled system are characterized by the cooperativity~\cite{grezes2015thesis},
\begin{equation}\label{eq:cooperativity}
  C = \frac{2g_\mathrm{ens}^2}{\kappa\Gamma},
\end{equation}
which compares the collective coupling rate of the ensemble to the dissipation rates of the resonator and the spins, quantifying how strongly the coherent coupling competes with these loss channels.
It is governed by three parameters: the inhomogeneous linewidth $\Gamma$, defined as the full-width at half-maximum (FWHM) of the ensemble spectral distribution; the cavity loss rate $\kappa$, the half-width at half-maximum (HWHM) of the resonator line; and the ensemble coupling strength $g_\mathrm{ens} = g_0\sqrt{N}$, where $g_0$ is the single-spin coupling strength and $N$ is the number of spins collectively coupled to the resonator.

When $C \gg 1$, the system is in the high-cooperativity regime, where the ensemble coupling strength dominates the dissipative response of the resonator and collective effects become prominent.
The high-cooperativity regime should be distinguished from the strong-coupling regime, where $g_\mathrm{ens} > \kappa,\,\Gamma$, so that the resonator and ensemble exchange excitations coherently, the normal-mode (ensemble Rabi) splitting is spectrally resolved, and coherent Rabi oscillations are observed.
In contrast, when $C < 1$, the system crosses into the low-cooperativity regime, where the spins have little influence on the resonator and collective effects are masked by dissipation.

When $C \approx 1$, the system reaches the ``impedance-matching'' condition, in which the coupled spin-ensemble--cavity system absorbs microwave tones arriving through the transmission line without reflection loss---the optimal working point of a quantum memory for itinerant microwave photons\cite{afzelius2013proposal}.
Accessing and tuning between these regimes therefore requires control of $\Gamma$, $\kappa$, or $g_\mathrm{ens}$.
However, these parameters are usually fixed by the device geometry and the spin concentration, and are thus difficult to adjust \textit{in situ} once the system is cold.
Therefore, reproducibly achieving the impedance-matching condition requires reliable \textit{in situ} control of at least one of these parameters, which would also enable a systematic study of the transition between the high- and low-cooperativity regimes, as well as the onset of strong coupling.
One established \textit{in situ} approach is to reshape the spectral distribution through spectral hole burning, which carves out the saturated spins to narrow the effective linewidth and suppress decoherence\cite{putz2014protecting, krimer2015hybrid, putz2017spectral}, but it can only narrow the distribution and cannot broaden it.

To this end, we developed an anti-Helmholtz coil integrated into the still shield of a dilution refrigerator.
This provides a tunable, \textit{in situ} technique to increase the width of the spectral distribution by applying a finely tuned constant gradient across the spin ensemble.

This paper is organized as follows.
In Sec.~\ref{sec:theory}, we calculate how a spin ensemble with constant spin concentration and initial inhomogeneous broadening is reshaped by an applied gradient.
In Sec.~\ref{sec:setup}, we present our device and measurement setup.
In Sec.~\ref{sec:results}, we reconstruct the spin-ensemble spectral distribution from the reflection spectrum to confirm our theory, and then use the device to probe the transition from the high- to the low-cooperativity regime: in continuous-wave measurements we demonstrate impedance matching, achieving near-perfect absorption by the spin ensemble of microwave tones incident from the transmission line, whereas in pulsed measurements we observe the disappearance of collective Rabi oscillations during the free induction decay (FID) as $\Gamma$ increases.


\section{Theory of Gradient Spectral Broadening}\label{sec:theory}
The Hamiltonian for a spin ensemble consisting of a large number of $N$ spins coupled to a single resonator mode is given by the Tavis-Cummings Hamiltonian\cite{tavis1968exact}.
\begin{equation}
\frac{\hat{H}_{TC}}{\hbar} = \omega_r \hat{a}^\dag \hat{a} + \sum_{j=1}^N \frac{1}{2}\omega_j\hat{\sigma}_{z,j} + \sum_{j=1}^N g_0(\hat{a}^\dag \hat{\sigma}_{-,j} +\hat{a}\hat{\sigma}_{+,j})
\end{equation}
where $\hbar$ is the reduced Planck's constant, $\omega_r$ is the resonator's angular frequency, $\hat{a}$ and $\hat{a}^\dag$ are the photon annihilation and creation operators, respectively, $\omega_j$ is the angular frequency of the spin at site $j$, $\hat{\sigma}_{z,j}$ is the Pauli-Z spin operator for the spin at site $j$, $g_0$ is the single spin-resonator coupling strength, and $\hat{\sigma}_{-,j}$, $\hat{\sigma}_{+,j}$ are the spin lowering and raising operators for the spin at site $j$, respectively.

Since we will be working with low excitation power, we make the Holstein-Primakoff approximation, replacing the Pauli operators with bosonic spin operators $\hat{\sigma}_{-,j}\rightarrow \hat{s}_j$, $\hat{\sigma}_{+,j}\rightarrow \hat{s}^\dag_j$, $\hat{\sigma}_{z,j}\rightarrow 2\hat{s}^\dag_j\hat{s}_j$, giving
\begin{equation}
\frac{\hat{H}_{TC}}{\hbar} = \omega_r \hat{a}^\dag \hat{a} + \sum_{j=1}^N \omega_j\hat{s}^\dag_j\hat{s}_j + \sum_{j=1}^N g_0(\hat{a}^\dag\hat{s}_j  +\hat{a}\hat{s}^\dag_j).
\end{equation}
To account for the inhomogeneous broadening, we must change these quantum operators to quantum field operators, letting $\sum_{j=1}^N\rightarrow N\int_{-\infty}^{\infty}\rho(\omega_s)\dd\omega_s$, $\hat{s}_j \rightarrow \hat{s}(\omega_s)$ and $\hat{s}_j^\dag \rightarrow \hat{s}^\dag(\omega_s)$, where $\rho(\omega_s)$ is the normalized spectral distribution, $\hat{s}(\omega_s)$ and $\hat{s}^\dag(\omega_s)$ are the annihilation and creation operators of the bosonic modes at frequency $\omega_s$ with commutation relation $[\hat{s}(\omega_s),\hat{s}^\dag(\omega_s')]=(N\rho(\omega_s))^{-1}\delta(\omega_s-\omega_s')$\cite{gardiner1985input}, and $\delta(\omega_s-\omega_s')$ is the Dirac delta function.
This results in
\begin{equation}
    \begin{aligned}
\frac{\hat{H}_{TC}}{\hbar} =& \omega_r \hat{a}^\dag \hat{a} + N\int_{-\infty}^{\infty}\omega_s\rho\hat{s}^\dag\hat{s}\dd\omega_s\\ +& gN\int_{-\infty}^{\infty}\rho [\hat{a}^\dag\hat{s}  +\hat{a}\hat{s}^\dag]\dd\omega_s,
    \end{aligned}
\end{equation}
where we have dropped the explicit dependencies on $\omega_s$ to avoid clutter.
The spectral distribution is characterized by the inhomogeneous linewidth $\Gamma$, which quantifies its spectral width, which is a function of the sample's intrinsic material properties.
When a constant magnetic field gradient is applied across the spin ensemble, the individual $\omega_j$ become a function of position along the direction of the gradient, modifying the shape of the spectral distribution.
We can produce this gradient using a pair of anti-Helmholtz coils. 

\begin{figure}[!ht]
  \centering
  \includegraphics[width=7cm]{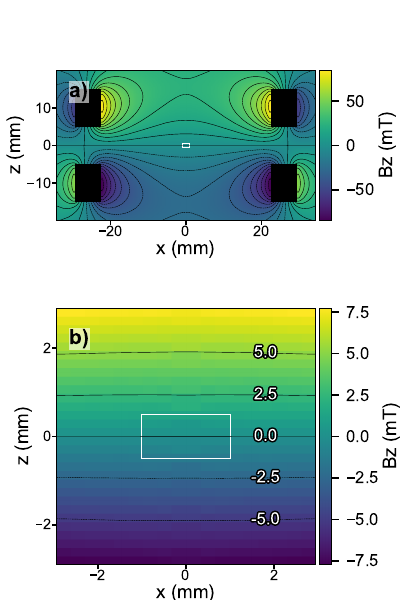}
  \caption{a) Cross section of the simulated static magnetic field $z$-component, produced by the anti-Helmholtz coils with the coil axis along the $z$-direction. The 4 black rectangles represent cross sections of the coils, the lines are contour lines of constant $B_z$, and the sample is contained in the white box outlined at the center of the plot. The simulation has 9000 Turns per coil and driven with a current of 200 mA. b) Close up of the sample region, where the field is shown to be well approximated as linear along the $z$-direction.}
  \label{fig1}
\end{figure}

In Fig.~\ref{fig1}a), we show the simulation results of such an anti-Helmholtz coil where the coil axis is along the $z$-direction.
The two black rectangles on the top and on the bottom are cross sections of the upper and lower coils, respectively.
To make this an anti-Helmholtz coil pair, the top and bottom coils are wound in opposite directions and connected in series.
In Fig.~\ref{fig1}b) we show a close up of the sample space, represented by the white rectangle in the center, demonstrating the linearly increasing magnetic field in the z-direction and constant field in the x-direction, in the sample space.

If we assume the initial spectral distribution, $\rho_0(\omega_s)$, to be uniform across the sample with initial inhomogeneous broadening $\Gamma_0$, we can represent the total spectral distribution, $\rho(\omega_s)$, as the convolution of $\rho_0(\omega_s)$ with the spectral distribution due to the applied magnetic field gradient $\rho_g(\omega_g)$.
\begin{equation}
  \rho(\omega_s) = \int_{-\infty}^\infty \dd \omega_g \rho_g(\omega_g) \rho_0(\omega_s-\omega_g).
\end{equation}
If the gradient is large such that the difference in transition frequency at either edge of the sample is much larger than $\Gamma_0$,
\begin{equation}\label{approximation}
\gamma\frac{\partial B_z}{\partial z}\Delta z\gg\Gamma_0,
\end{equation}
where $\gamma$ is the gyromagnetic ratio of the spins, $z$ is the coordinate along the magnetic field gradient, and $\Delta z$ is the maximum width of the spin ensemble along the z-direction.
Then $\rho_0(\omega)$ is well approximated as a Dirac delta function and the integral reduces to 
\begin{equation}\label{density theory}
  \rho(\omega_s) \approx \rho_g(\omega_s).
\end{equation}
\section{Experimental setup}\label{sec:setup}
\begin{figure}[!ht]
  \centering
  \includegraphics[width=8.5cm]{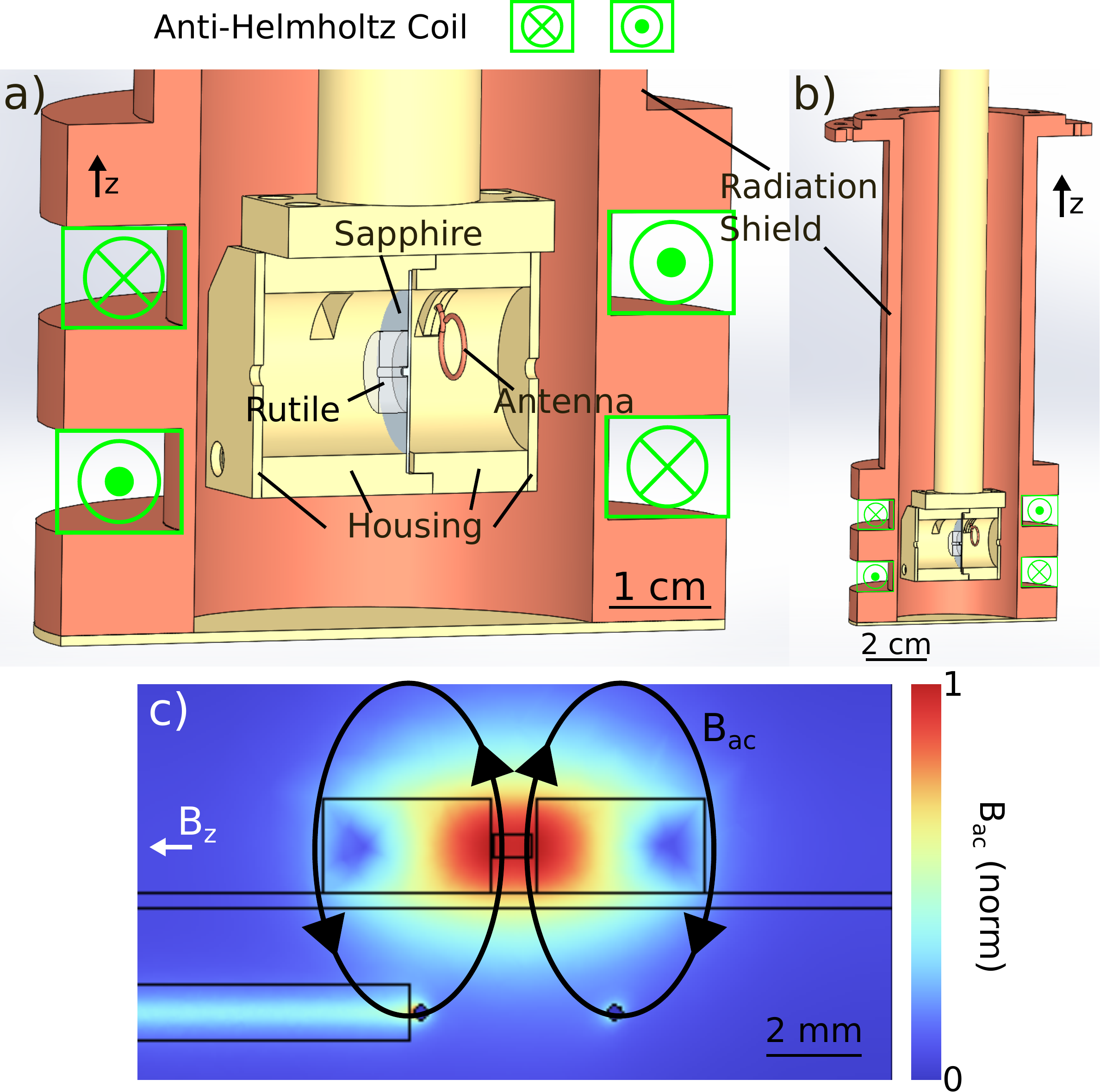}
  \caption{(a) Anti-Helmholtz coils integrated into the tail of the still shield, inside which the dielectric resonator is housed. 
  (b) The entire still shield tail and the device, which sit inside the superconducting vector magnet (not shown here, see the supplementary material). 
  (c) Simulated ac magnetic-field amplitude of the resonator's fundamental mode. The direction of the ac field is indicated by the black arrowed circles, and the static magnetic field by $B_z$. The rutile crystal has an outer diameter of 8~mm, an inner diameter of 1~mm, and a height of 2~mm.}
  \label{fig2}
\end{figure}

We used a three-dimensional dielectric resonator consisting of a cylindrical rutile (TiO$_2$) crystal enclosed in a copper housing\cite{probst2014three}, as shown in Fig.~\ref{fig2}a).
Bringing the electron spins into resonance with the magnetic field of the fundamental TE$_{01\delta}$ mode at $4.557$~GHz requires a static magnetic field of $\sim163$~mT applied across the sample.
In this field range, superconducting resonators are degraded by field-induced losses, whereas a dielectric resonator in a copper enclosure is insensitive to such fields.
At low temperature, rutile has a high relative permittivity ($\epsilon_\perp \approx 112$\cite{tobar1998anisotropic}) and a low dielectric loss, producing a compact mode with a measured internal quality factor of $Q_\mathrm{int} \gtrsim 10^{5}$ and a uniform coupling strength throughout the sample space\cite{kato2023high,hamamoto2024dielectric}.

We simulated the fundamental mode (Fig.~\ref{fig2}c) to estimate its resonance frequency, $4.67$~GHz, and single-spin coupling strength, $g_0 = 82\,\mathrm{mHz}$; the measured value of $4.557$~GHz agrees to within $2.5\%$.
Incoming microwave tones are delivered to the resonator through a loop antenna, whose position is adjusted before cool-down to set the external quality factor to $Q_\mathrm{ext} \approx 1000$ (Fig.~\ref{fig2}a).
We then characterized the resonator by measuring its reflection spectrum through this antenna with a vector network analyzer (VNA); the microwave setup is described in detail in the supplementary material.

To create a magnetic field gradient and thereby control the inhomogeneous broadening, the anti-Helmholtz coil consists of a pair of coils, each with 9000 turns of 102~\unit{\micro\meter} NbTi wire, mounted concentrically with the sample space of the dielectric resonator (Fig.~\ref{fig2}a,b).
The centers of the anti-Helmholtz coil and the dielectric resonator also coincide with that of the superconducting three-dimensional vector magnet, which provides a uniform static field to control the electron spin resonance (ESR) frequency.
Further details of the coil mounts and wiring are provided in the supplementary material.

The spin ensemble consists of the unpaired electrons of 2,2-diphenyl-1-picrylhydrazyl (DPPH) molecules placed inside the rutile cylinder (Fig.~\ref{fig2}c).
DPPH is paramagnetic above $\sim 4$~K, while antiferromagnetic correlations become significant at lower temperatures \cite{fujito1981magnetic, mergenthaler2017strong}.
Cooling to millikelvin temperatures would therefore complicate the analysis due to these antiferromagnetic correlations. 
We thus maintained the dilution refrigerator mixing chamber at 5.5~K, where the ensemble remains paramagnetic with the number of paramagnetic spins on the order of $10^{15}$ (see Supplementary Material).

\section{Results}\label{sec:results}
To determine the spin spectral distribution, we measured the reflection spectrum by probing the spin-resonator system through the antenna, using the VNA (supplementary material).
We then fit the data to the following equation \cite{diniz2011strongly}:
\begin{equation}\label{eq:reflection}
  r(\omega) = \frac{i\kappa}{(\omega -\omega_r) +i\kappa/2 - K(\omega)}-1,
\end{equation}
where $K(\omega)$ is the linear response function that contains all the spectral information about the spin ensemble, $\kappa$ is the cavity loss rate in the limit where the internal cavity loss is negligible compared to the the external loss and $i$ is the imaginary unit.
The function $K(\omega)$ can be calculated applying the Sokhotski-Plemelj integral to the spectral distribution.
Conversely, after solving Eq.~\ref{eq:reflection} for $K(\omega)$, the spectral distribution can be obtained from \cite{kurucz2011spectroscopic}
\begin{equation}\label{eq:density measured}
  \rho (\omega) = -\frac{1}{\pi}\im K(\omega).
\end{equation}

Practically, this method only works for the section of $\rho(\omega)$ within the bandwidth of the resonator-spin ensemble system because outside of the resonator bandwidth, $r\approx 1$ regardless of $K(\omega)$.
To obtain the $\rho(\omega)$ when $\Gamma$ is larger than $\kappa$, we measure different sections of the spectral distribution by changing $B_0$ and holding the gradient constant. We then reconstruct the full distribution by piecing these sections together.

\subsection{Ensemble Spectral Distribution}
\begin{figure}[!ht]
  \centering
  \includegraphics[width=8.5cm]{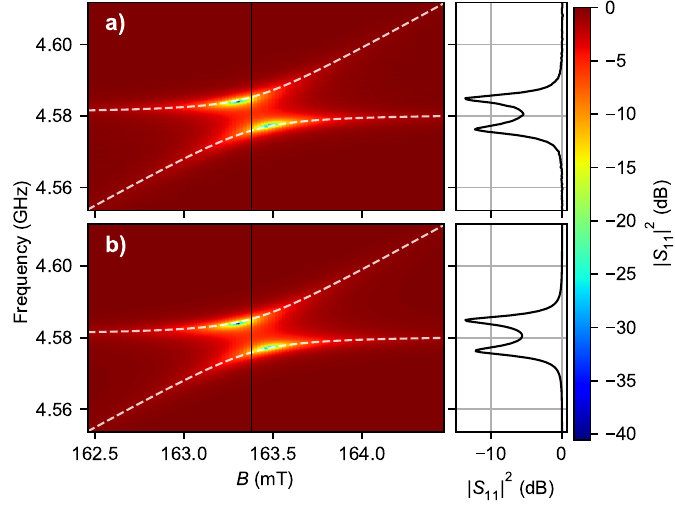}
  \caption{Reflection spectrum as a function of probe frequency and magnetic field when no magnetic field gradient was applied, corresponding to the high cooperativity regime. a) is the experimental data and b) is the fitted spectrum using the fitted parameters from a). The white dashed lines are the resonance frequencies of the polaritons and the spectra on the right correspond to black vertical lines, where the spins are on resonance with the resonator.}
  \label{fig3}
\end{figure}

\begin{figure}[!ht]
  \centering
  \includegraphics[width=8.5cm]{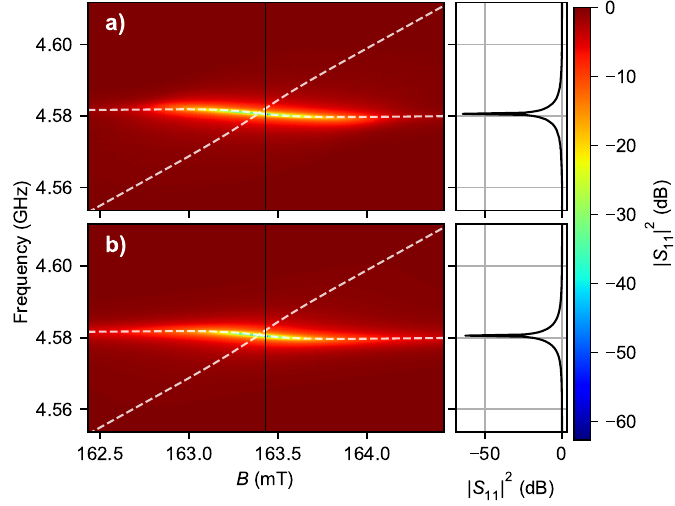}
  \caption{Reflection spectrum as a function of probe frequency and magnetic field when a magnetic field gradient is applied such that $C=1$. a) is the experimental data and b) is the fitted spectrum using the fitted parameters from a). The white dashed lines are the resonance frequencies of the polaritons and the spectra on the right correspond to black vertical lines, where the spins are in resonance with the resonator.}
  \label{fig4}
\end{figure}

The measured and fitted reflection coefficient function when no magnetic field gradient is applied are plotted in Fig.~\ref{fig3} a) and b), respectively and likewise with an applied magnetic field gradient in Fig.~\ref{fig4} a) and b).
To the right of every spectrogram, we included the spectrum for the case where the spins are on resonance with the resonator.
By fitting the data to Eq.~\ref{eq:reflection}, and assuming a Lorentzian distribution for the spin ensemble spectral distribution, we obtained the values of $g_{ens}/2\pi=5.3$~MHz, $\kappa/2\pi = 2.3$~MHz and $\Gamma_0/2\pi= 7.9$~MHz as fitting parameters.
The magnetic field gradient reduces the cooperativity from $C=3.2$ (Fig.~\ref{fig3}), where the level anti-crossing (LAC) is visible, to $C=1$ (Fig.~\ref{fig4}), where the LAC collapses into a single dip that falls below the noise floor.

\begin{figure}[!ht]
  \centering
  \includegraphics[width=8.5cm]{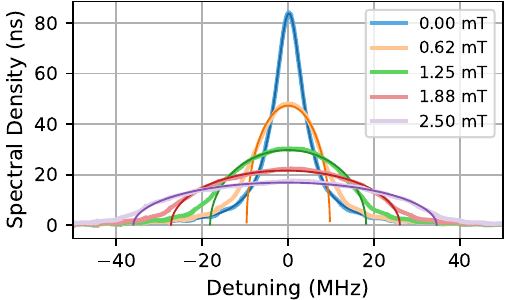}
  \caption{Spin spectral distribution data(light) and fit(dark) for various magnetic field gradients. The legend lists $\Delta B$, the magnetic field difference on either side of the sample. The zero current data is fit to a Lorentzian function, while all the others are fit to semi-ellipses using only the data in the central region.}
  \label{fig5}
\end{figure}

We measured similar spectra for 5 different magnetic field gradients, corresponding to currents ranging from 0 to 200 mA.
Using Eq.~\ref{eq:reflection} and Eq.~\ref{eq:density measured}, we calculated $\rho$ and plotted the result in Fig.~\ref{fig5} (lighter traces).
Without any applied gradient the distribution is fitted to a Lorentzian distribution (dark blue).

To calculate the functional form of $\rho_g(\omega)$, we start by calculating $\rho_s(z)$, the spin concentration as a function of the z-direction, we assume a constant spin density and integrate over the 2 directions of the cylindrical sample space with no gradient
\begin{equation}
  \rho_s(z) = c\int_0^{x_0}\dd x \int_{-\sqrt{r^2-z^2}}^{\sqrt{r^2-z^2}}\dd y= 2c{x_0}\sqrt{r^2-z^2},
\end{equation}
where $c$ is the spin concentration and $r$ is the radius of the sample. 
Making a change of variables to the spin transition frequency,
\begin{equation}
  z \rightarrow\omega = 2\pi\gamma\left(B_0 + \frac{\dd B}{\dd z} z\right),
\end{equation}
we obtain a semi-ellipse, shifted to the central transition frequency of the spin ensemble
\begin{equation}
  \rho_g(\omega) = cx_0\sqrt{r^2-\left[\frac{\omega/2\pi - \gamma B_0}{\gamma(\dd B/\dd z)} \right]^2}.
\end{equation}
We fit the measured spectral distribution to this function and find good agreement when $\rho_g(\omega)$ is much larger than $\Gamma_0$, as expected from the inequality Eq.~\ref{approximation}.
For nonzero gradients that do not fulfill this condition, we find a good fit in the central region, but divergence at either side of $\rho(\omega)$.

From this fit and the known size of the sample space, we can characterize the gradient generated by the anti-Helmholtz coils as $12.5$~mT$\cdot$mm$^{-1}$A$^{-1}$, corresponding to $350$~MHz$\cdot$mm$^{-1}$A$^{-1}$ for DPPH, which agrees with the simulation in Fig.~\ref{fig1} within $3.8 \%$.
The legends in the figures list the magnetic field difference on either edge of the sample space ($\Delta B = 12.5 $ mT mm$^{-1}$A$^{-1}\times 1$ mm $\times I$).

\subsection{Frequency Domain: Effective Impedance Matching}
\begin{figure}[!ht]
  \centering
  \includegraphics[width=8.5cm]{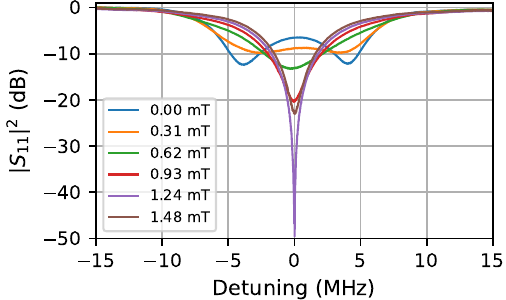}
  \caption{Reflection spectra of the resonator-spin ensemble system for different applied magnetic field gradients. The legend indicates $\Delta B$, the magnetic field difference on either edge of the sample.}
  \label{fig6}
\end{figure}
Impedance matching the cavity-spin ensemble system to a transmission line is important to maximize the absorption of incoming photons.
It has been shown that for a Lorentzian inhomogeneous spectral distribution, this is achieved under two conditions, the first when $C=1$ and zero detuning, and the second, if $C\gg 1$, happens when $\Gamma=\kappa$ when driving on resonance with one of the polaritons\cite{afzelius2013proposal}.
We note that both these condition depend on $\Gamma$, implying that our device can be used to tune the effective impedance and maximize the absorption in both cases.
We demonstrate this for the case of $C=1$, where we measure the spectra for a series of magnetic field gradients while keeping the central frequency of the spin ensemble on resonance with the resonator.
We show in Fig.~\ref{fig6} that when no gradient is present, we observe two dips corresponding to the two polaritons, as shown in Fig.~\ref{fig3} at 163.3~mT.
By applying a small gradient such that the difference in magnetic field across the 1~mm diameter sample is $\Delta B = 0.62$~mT, these polaritons transition to a single dip with a $14$~MHz bandwidth. 
Further increasing the gradient, the dip narrows and deepens until $\Delta B= 1.24$~mT where the effective impedance is matched, $C=1$, corresponding to $99.999\%$ absorption by the ensemble\cite{afzelius2013proposal}.
Increasing the gradient beyond this value, the system enters the low cooperativity regime, where the absorption on resonance decreases.

\subsection{Time Domain: Coherent Rabi Oscillations}
\begin{figure}[!ht]
  \centering
  \includegraphics[width=8.5cm]{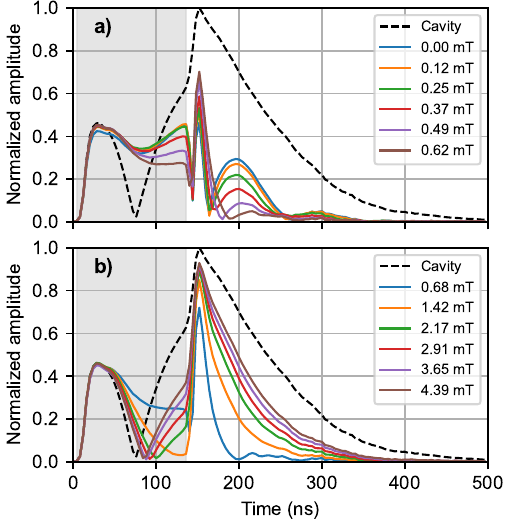}
  \caption{Experimental data of the FID for different magnetic field gradients across the sample, separated over two plots for clarity. a) shows the high cooperativity regime, where collective Rabi oscillations between cavity and spin ensemble are visible and, b) shows the FID in the low cooperativity regime, where the oscillations become too small to observe.}
  \label{fig7}
\end{figure}

\begin{figure}[!ht]
  \centering
  \includegraphics[width=8.5cm]{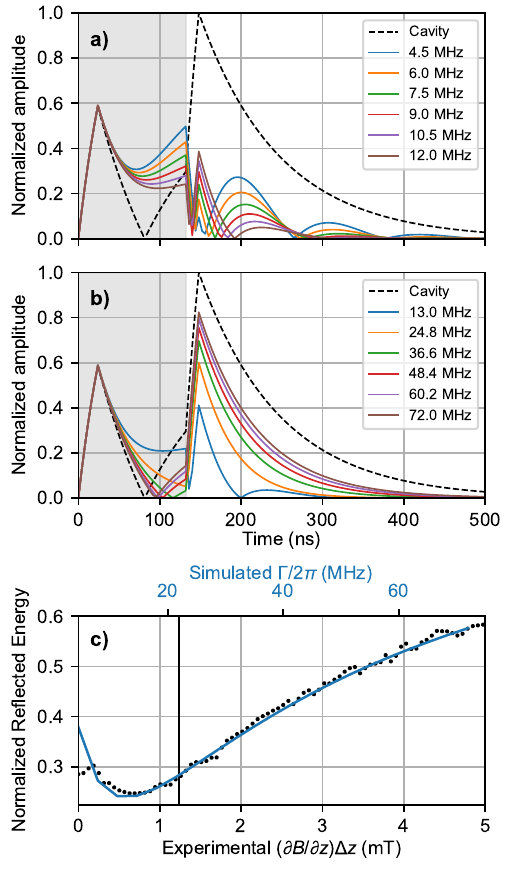}
  \caption{Simulation of the FID for different spin spectral broadening across the sample. a) The FID in the high-cooperativity regime, where collective Rabi oscillations between cavity and spin ensemble are visible. b) The FID in the low-cooperativity regime, where the oscillations become too small to observe. c) The integral of amplitude squared, normalized by the integral of the bare-cavity ring-down amplitude squared. The black dots correspond to the experimental data from Fig.~\ref{fig7}, as a function of the field difference on either end of the sample (bottom axis), and the blue line corresponds to the simulations from a) and b) as a function of the dephasing rate (top axis). 
  The vertical solid black line indicates the $C=1$ impedance-matching point obtained from the continuous-wave measurement.}
  \label{fig8}
\end{figure}

To show this transition in the time domain, we applied a 132 ns square pulse to the spin ensemble through the antenna with the spins on resonance with the resonator and measured the FID for a series of magnetic field gradients.
A detailed description of the pulse generation and time domain measurement setup can be found in the supplementary material.
These pulses and the following FID field amplitudes are shown in Fig.~\ref{fig7}a) and b) where the time the pulses were applied are highlighted in gray.
The 132 ns square pulses were chosen in order to simultaneously address both polaritons when no gradient is applied, corresponding to the maximum splitting.
For clarity, we plotted the FID amplitudes where the ensemble Rabi oscillations are visible in Fig.~\ref{fig7}a) and those where the ensemble Rabi oscillations are not visible in Fig.~\ref{fig7}b).
We include the bare cavity ring-down measurement as a reference.

To confirm these results we simulated the FID signal and demonstrate the transition from the strong to the weak coupling regime.
In the simulation, we model the spin ensemble, using the Holstein-Primakoff approximation, as a harmonic oscillator which is coupled to a second harmonic oscillator representing the resonator.
A more realistic method to simulate the FID signal, would be to split the simulation into many sub-ensembles and track how these sub-ensemble lead to dephasing.
However, since we do not need to track the phase after the sub-ensembles dephase inhomogeneously and since $\kappa<\Gamma/2$,  the cavity filters a portion of the ensemble.
This leads to dephasing that can be approximated as Markovian.
We therefore model the dephasing using a dissipative superoperator with $\sqrt{\Gamma/2} \hat{s}$ in a Lindblad master equation.
Details about the simulation can be found in the supplementary material.
We again split data, showing the FID signals where coherent oscillations are visible in Fig.~\ref{fig8}a) and the weak coupling regime where the coherent oscillations are not observable in Fig.~\ref{fig8}b).

In Fig.~\ref{fig8}c) we show the reflected pulse energy, obtained by integrating the squared magnitude of the measured signal over time and normalizing it by the same integral for the bare-cavity ring-down, which involves no spin absorption.
Because $\kappa_\mathrm{int}\ll\kappa_\mathrm{ext}$, internal losses are negligible, so any energy not absorbed by the spin ensemble returns to the measurement line, either promptly reflected or subsequently leaked back out from the resonator.
The reduction in the reflected energy relative to the bare-cavity reference therefore corresponds to the energy absorbed by the ensemble.
We see that the maximum absorption occurs neither at the same magnetic field gradient nor at the same level as in the frequency domain (Fig.~\ref{fig6}).
In the continuous-wave measurement, we achieved maximum absorption at $\Delta B = 1.24$~mT, corresponding to $50$~dB absorption with a probe bandwidth of $1$~kHz.
Because the short pulse has a $6.71$~MHz bandwidth, much of its energy is reflected, and the maximum absorption is only $6$~dB, reached at $\Delta B = 0.69$~mT.
At this gradient the cooperativity is $C=1.8$ rather than $1$ (Fig.~\ref{fig6}), where the reflection dip is significantly wider than at $C=1$.

In order to efficiently absorb itinerant microwave photons, the device bandwidth must be compatible with the duration of these photons, typically $\sim100$~ns\cite{pechal2014microwave}.
The bandwidth of the present device at the impedance-matching point ($C=1$) is not yet wide enough to accommodate such pulses.
Since impedance matching occurs at $\Gamma = 2g_\mathrm{ens}^2/\kappa$, widening this bandwidth requires a larger ensemble coupling $g_\mathrm{ens} = g_0\sqrt{N}$.
In practice, increasing the spin concentration is the most effective route to a larger $g_\mathrm{ens}$, and hence to a wider bandwidth. 
In addition, we can widen the sample space somewhat, since a significant fraction of the ac magnetic field still resides within the rutile (Fig.~\ref{fig2}c)), while keeping the resonator frequency and the ac-field homogeneity approximately unchanged.
For example, NV centers in diamond at a concentration of 15~ppm would provide a total of $N \sim 10^{16}$ spins, giving $g_\mathrm{ens} \sim 10\,\mathrm{MHz}$ and hence a device bandwidth compatible with typical itinerant microwave photons.

\section{Conclusion}

We demonstrate impedance matching between the transmission line and a coupled spin-ensemble--cavity system using a magnetic field gradient generated by an anti-Helmholtz coil.
We further show that this device enables \textit{in situ} tuning of the inhomogeneous linewidth, driving the coupled system from the strong- to the weak-coupling regime. 
This platform provides an \textit{in situ} testbed for studying how collective effects, such as superradiance, subradiance, and pulsed echo trains, evolve across the transition from the strong- to the weak-coupling regime. 
Furthermore, it enables modification of the full spectral distribution by tailoring the physical shape of the spin-ensemble host material.


\begin{acknowledgments}
We thank \c{C}.~O. Girit, M. Stern, and P. Bertet for helpful discussions, and Y. Feng for providing access to the dilution refrigerator. 
We are also grateful to the members of the Hybrid Quantum Device Team within the Science and Technology Group and the Experimental Quantum Information Physics Unit at Okinawa Institute of Science and Technology (OIST) Graduate University for their insights. 
We acknowledge the support of the Engineering Section of OIST for machining some parts. 
This work has been supported by the JST Moonshot R\&D Program (Grant No.~JPMJMS2066), JST PRESTO (Grant No.~JPMJPR15P7), JSPS KAKENHI Grant-in-Aid for Scientific Research (B) (Grant No.~18H01817), Grant-in-Aid for Scientific Research on Innovative Areas (Grant No.~18H04295), the Nakajima Foundation, the Sumitomo Foundation, the Research Encouragement Grant from the AGC Inc.\ Research Foundation, and OIST Graduate University.
\end{acknowledgments}

\section*{Data Availability Statement}
The data that support the findings of this study are available from the corresponding author upon reasonable request.


\clearpage
\onecolumngrid
\begin{center}
  \textbf{\large Supplemental Material}\\[3pt]
  \textbf{Controlling the Inhomogeneous Broadening and Impedance Matching of a Spin Ensemble}
\end{center}
\vspace{6pt}
\twocolumngrid

\setcounter{figure}{0}
\renewcommand{\thefigure}{S\arabic{figure}}
\setcounter{equation}{0}
\renewcommand{\theequation}{S\arabic{equation}}
\setcounter{section}{0}
\renewcommand{\thesection}{S\arabic{section}}

\section{Microwave Control Setup}
\begin{figure*}[!h]
  \centering
  \includegraphics[width=15cm]{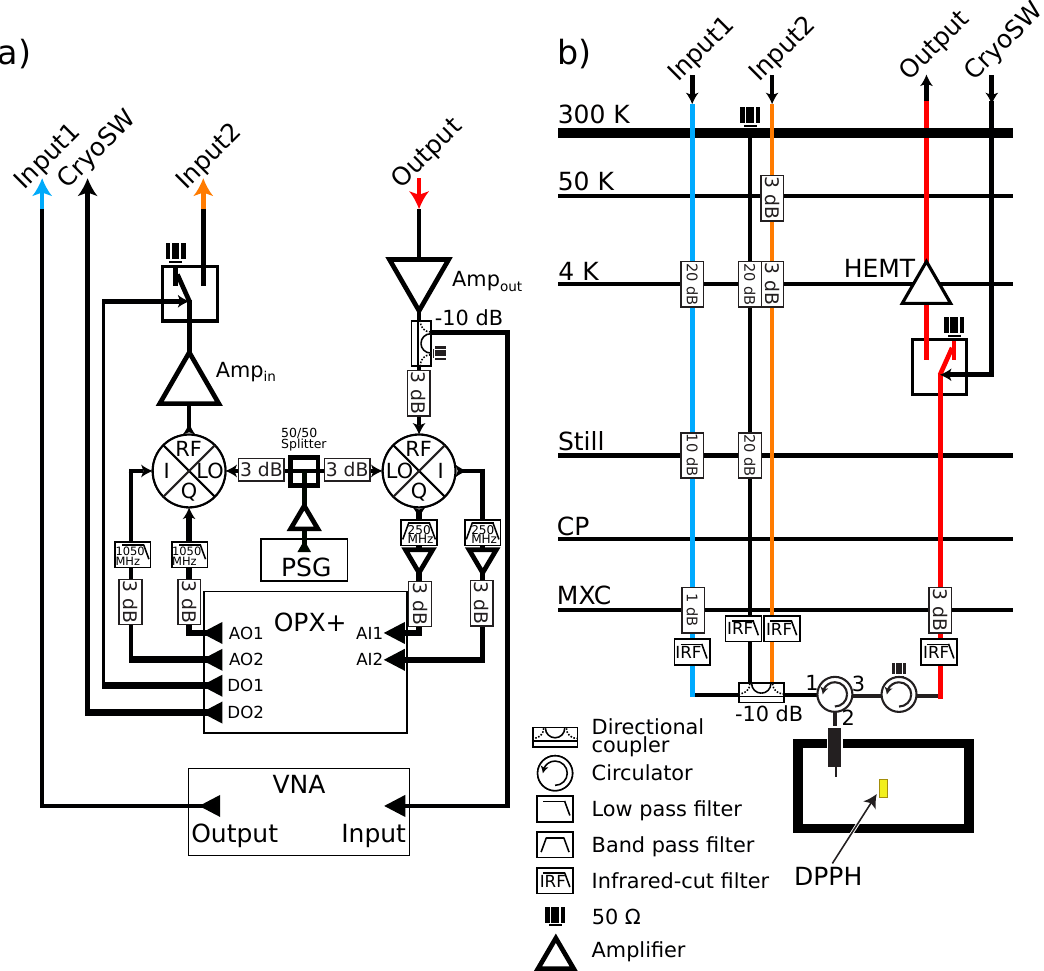}
  \caption{Microwave control schematic. a) Room temperature circuit. The spectrum is obtained using a \acrshort{vna} that scans a frequency range with a weak tone through Input 1 of the fridge. The signal then returns from the fridge, is amplified and a small portion is sent back to the \acrshort{vna}. The power sent to the \acrshort{iq} mixer is ignored. For the pulse experiments, a \acrshort{mw} tone is constantly sent to the \acrshort{lo} ports of the two \acrshort{iq} mixers from the PSG. The OPX+ generates \gls{rf} pulses which are mixed with the \acrshort{lo}, sending a \acrshort{mw} pulse to Input 2 of the fridge. The signal returns through the Output and is mixed back down to the \acrshort{rf} frequency that the OPX+ can digitize through its analog inputs. b) The cryogenic circuit. The \gls{cw} tones enter through Input 1 and pass through thermalized attenuators designed to reduce the thermal photons guided to the colder stages. The pulses enter the fridge through Input 2 and follow a similar path to the \acrshort{cw} tones except the directional coupler is used for thermal photon attenuation instead of an attenuator at the mixing chamber stage. Signals coming from the resonator are separated from the input signals by the circulator and sent up the amplification chain. The cryogenic switch just before the \acrshort{hemt} protects the \acrshort{hemt} from the high power pulses coming from Input 2. The warmest three stages are labeled by their temperatures and the coldest stages are the \gls{still}, \gls{cp} and \gls{mxc} stages.}
  \label{fig:mw_setup}
\end{figure*}

A schematic of the room-temperature and cryogenic microwave setups is shown in Fig.~\ref{fig:mw_setup}a) and b), respectively.
In a), the probe microwave tone from a \gls{vna} scans a range of frequencies near the cavity resonance with low power going from the output of the \gls{vna} to Input 1 of the fridge (blue line).
After entering the fridge, shown in b), the signal passes through a series of thermalized attenuators, which reduce the flow of thermal photons through the transmission lines. The signal then passes through a directional coupler, losing only $10\%$ of its power before entering port 1 of a circulator that guides the signal to the cavity-spin ensemble system.
The reflected signal returns to port 2 of the circulator, then to the measurement line (red), where it is separated from the incident signal and amplified by a \gls{hemt} amplifier.
Before the HEMT, a cryogenic microwave switch is installed but it remains on throughout this experiment.
At room temperature, shown in a), the signal is amplified again and 10$\%$ is sent back to the \gls{vna} that measures and records the data.
We use in-house written Python drivers\cite{couillard_keysight_E5071C} and scripts\cite{couillard_hqd_epro}. 

Once the spin spectrum has been obtained and the resonant magnetic field identified, we carry out pulse measurements. 
A vector microwave generator (Keysight PSG E8247C) sends a constant \gls{mw} tone with its power split equally for up and down conversion.
The frequency of this \gls{mw} tone is set to 250~MHz less than the resonance frequency, which ensures any \gls{lo} leaking from the \gls{iq} mixer will be far off resonance with the spins.
The pulses are generated with the Quantum Machines OPX+ and come out through the \gls{ao}s labeled AO1 and AO2.
These outputs send pulses centered at the \gls{if} of 250~MHz which are sent to the \gls{iq} mixer for \gls{ssb} mixing with the \gls{mw} signal in order to shift all the \gls{mw} power to the cavity-spin ensemble resonance frequency.
The pulse generated after mixing is amplified by Amp\textsubscript{in} to 3 W using a Mini-circuits ZVE-3W-83+.
During the time the pulse passes through the switch following Amp\textsubscript{in}, the \gls{do}, labeled \gls{do}1 of the OPX+ closes the circuit to send the signal to Input 2.
When no pulse is being sent, the switch is set to send any background noise from Amp\textsubscript{in} to a 50 $\Omega$ load resistor, preventing it from entering the fridge.
The pulse enters the fridge and passes through thermalized attenuators, as with Input 1, with the main difference being that only 10\% of the signal goes through the directional coupler toward the sample.
This is used instead of a thermalization attenuator because the high power of these pulses would warm up the \gls{mxc} stage.
This way it sends the signal, with thermal photons, back up to warmer stages that have more cooling power.
The signal that is reflected off the cavity is sent up through the output line but is now attenuated by 40 dB at the cryogenic switch to protect the \gls{hemt}.
Once back at room temperature, the signal is amplified again, mixed back down to the \gls{if}, and collected by the OPX+ through the \gls{ai}s labeled AI1 and AI2.

\section{Anti-Helmholtz Coil Mounting and Wiring}
\begin{figure}[!h]
  \centering
  \includegraphics[width=8.5cm]{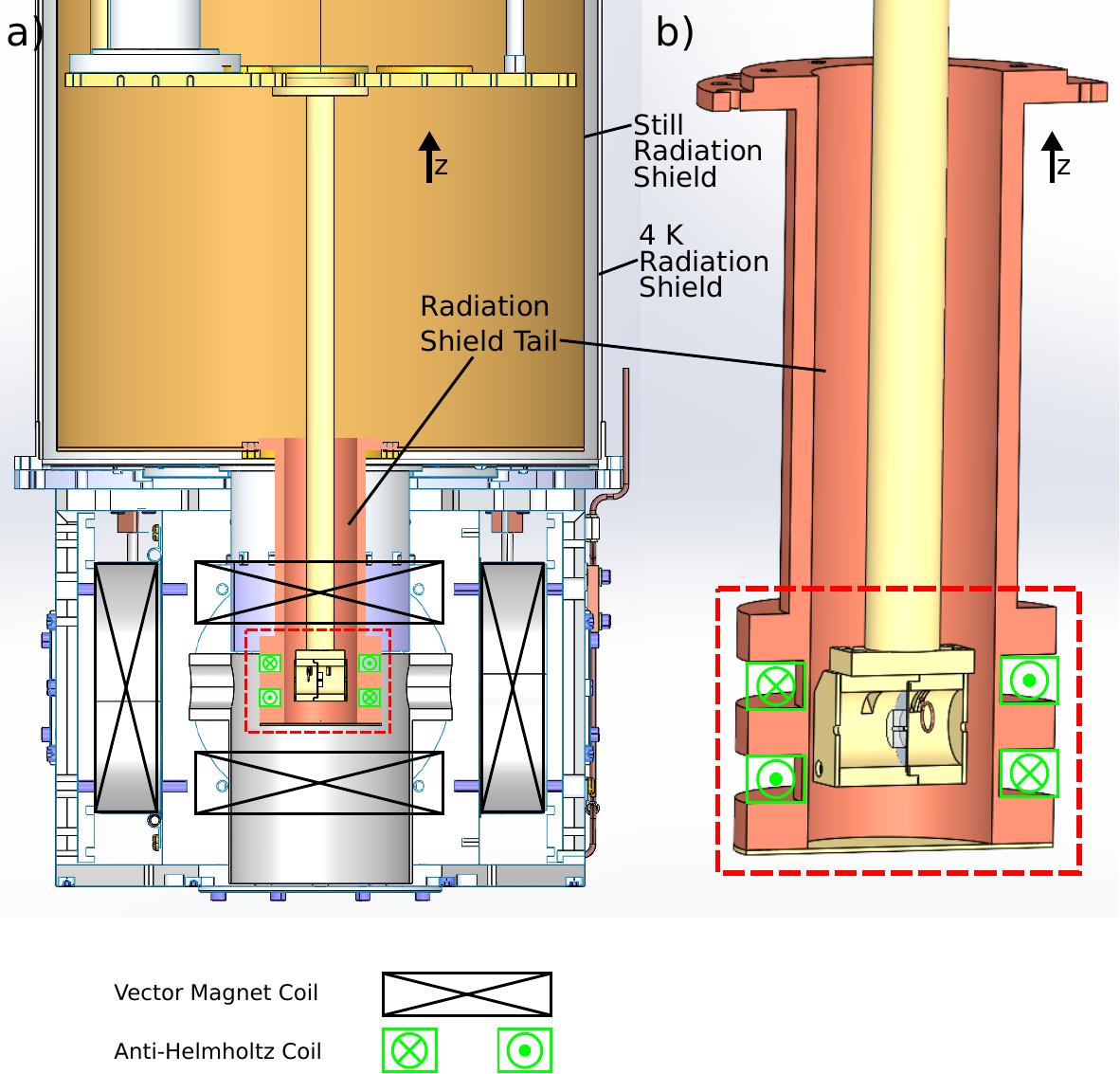}
  \caption{\gls{cad} file of the cryogenic setup. In a) we see how the resonator is thermalized to the \acrshort{mxc} plate and centered with the vector magnet. The vector magnet is attached to the radiation shield of the 4 K stage and the \acrshort{still}-shield fits just on the inside of the 4 K shield. The \acrshort{still}-shield contours the magnet and the custom, anti-Helmholtz, tail section goes in the bore of the vector magnet. In b) we have removed the fridge and enlarged the parts we fabricated: the anti-Helmholtz tail, the resonator and the thermalization rod. Image courtesy of Bluefors.}
  \label{fig:still_shield_mag}
\end{figure}
To support the coils, we integrated the coil mounts in a custom tail for the \gls{still} radiation-shield of the dilution refrigerator.
This tail section is what allows the radiation shield to fit between the sample space, which is thermalized to the \gls{mxc} stage, and the fridge's superconducting magnet, which is thermalized at the 4 K stage.
We show \gls{cad} files of the custom tail in the magnet in Fig.~\ref{fig:still_shield_mag}a).
The custom tail is shown with the resonator and thermalization rod in Fig.~\ref{fig:still_shield_mag}b) without the rest of the fridge which highlights the coil mounts and how they align with the resonator and sample.

Machining of the tail and winding of the coil around the mounts were outsourced to a machining company, which fabricated it from oxygen-free copper that was subsequently gold plated.
The wire used was Supercon's 54S43, consisting of 54 filaments of NbTi embedded in copper of $79~\mu$m in diameter and coated by an insulator that increases the diameter to $102~\mu$m\cite{Supercon54S43Spec}.
\begin{figure}[]
  \centering
  \includegraphics[width=8.5cm]{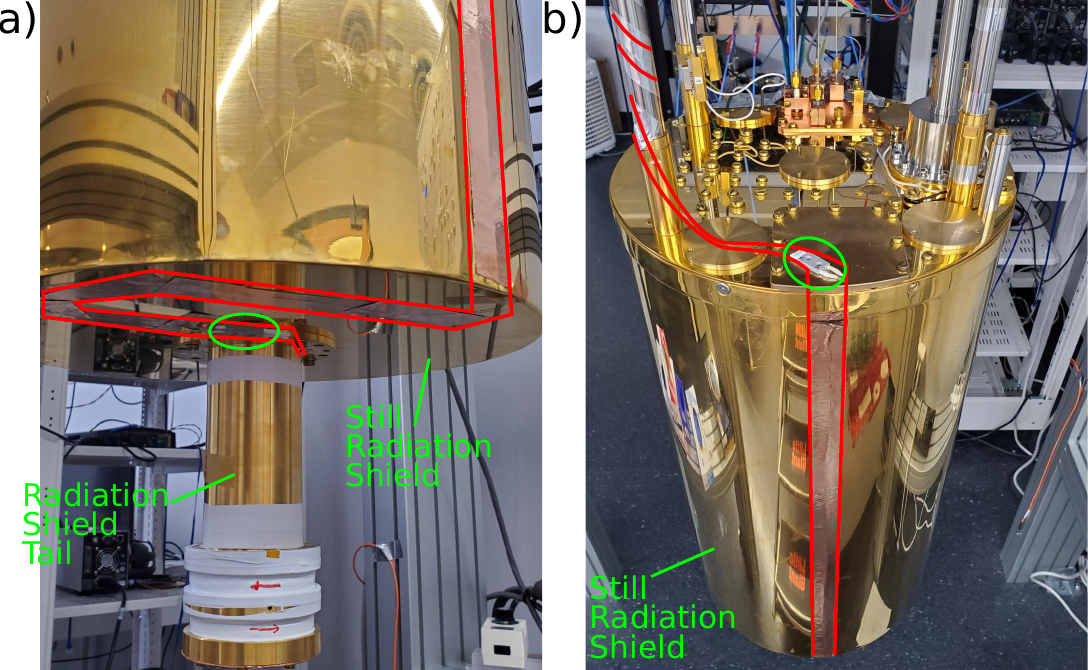}
  \caption{Pictures of the wiring used to drive the current in the anti-Helmholtz coils. The red lines are drawn on either side of the wires and the copper tape fixing the wires to the shield. The green ovals show the position of 2-prong connectors used to detach the wire when removing the \acrshort{still}-shield from the fridge or the tail from the \acrshort{still}-shield. a) The lower section where the \acrshort{still}-shield meets the tail. b) The upper section where the \acrshort{still}-shield meets the distillation stage plate. The wire is taped to the post on the left and thermalized with a bobbin at the 4 K stage before being soldered to a 25-pin micro-D connector (not in picture).}
  \label{fig:wire_and_connector}
\end{figure}

Since the coil is mounted on the exterior of the radiation shield, the wiring is taped to the exterior of the \gls{still}-shield using copper tape and connected to the 25-pin micro-D connector as shown in Fig.~\ref{fig:wire_and_connector}.
The coils were driven using two Yokogawa GS200 units as current sources, connected in parallel, that ramped the current using an in-house Python script to protect the GS200 from sinking the current from the high-inductance load coils during ramp-down.

\section{Total Number of Paramagnetic Spins}
As the temperature decreases, the magnetic susceptibility of \gls{dpph} starts to decrease significantly around 10 K\cite{Grobet1978spin} and continues until it reaches its antiferromagnetic phase transition temperature of 0.3 K\cite{prokhorov1963antiferromagnetism}.
Between these two temperatures, the magnetic susceptibility decreases due to pairing of electrons, removing their contribution to the \gls{epr} signal.
To calculate the number of spins at these temperatures, we use the Bleaney--Bowers model\cite{bleaney1952anomalous}, which treats the spins as dimers that occupy a singlet ground state and become paramagnetic only when thermally excited to the triplet state.
The population ratio can then be calculated from the Boltzmann distribution,
\begin{equation}\label{eq:paramag_fraction}
  n_\mathrm{BB} = \frac{3e^{-\frac{T_c}{T}}}{1+3e^{-\frac{T_c}{T}}},
\end{equation}
where $T_c$ is the characteristic temperature measured as 17.6 K for \gls{dpph}\cite{fujito1981magnetic}, and $T$ is the sample temperature which was 5.5 K for our experiments.

\begin{figure}[!h]
  \centering
  \includegraphics[width=8.5cm]{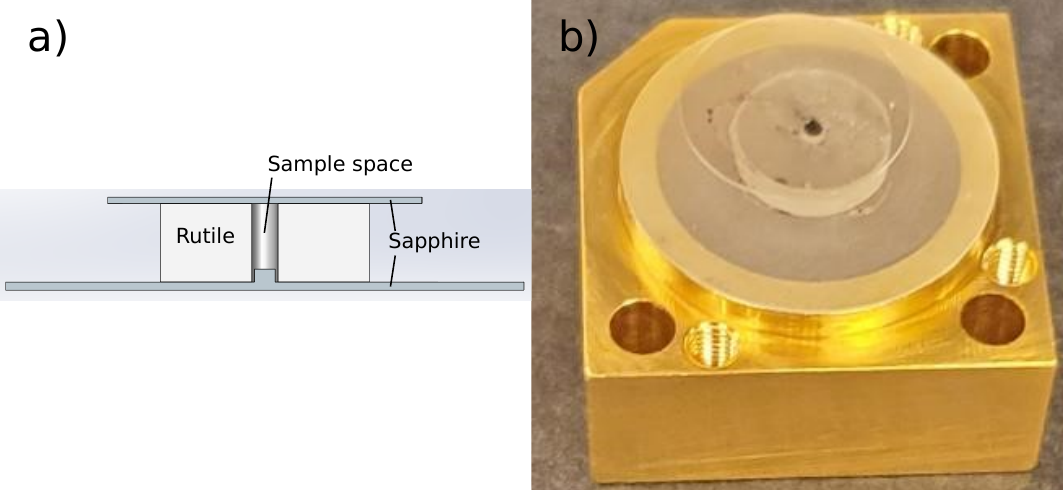}
  \caption{Rutile resonator with sapphire on top and below that are both held with vacuum grease in order to prevent the \acrshort{dpph} from falling due to gravity. The resonator was placed sideways in the refrigerator. a) A \acrshort{cad} rendering of the rutile and sapphire. b) A picture of the rutile and sapphire with \acrshort{dpph} in the sample space and placed on the lower half of the resonator housing.}
  \label{fig:dpph_sample_space}
\end{figure}

The \gls{dpph} is held in by closing off the top of the sample space with a small piece of sapphire to ensure the \gls{dpph} does not fall out due to gravity and vibrations when the resonator is placed in the dilution fridge.
Both pieces of sapphire are held with a small dab of vacuum grease as shown in Fig.~\ref{fig:dpph_sample_space}.

To maximize the coupling strength, we fill the sample space to maximum capacity.
To estimate the number of molecules, we measure the volume of the sample space and use the known density of \gls{dpph}.
Using the volume of the sample space, a mass density of $1.485$ mg$/$mm$^3$, mass of $6.55\times 10^{-19}$ mg$/$molecule, and assuming a minimum $36\%$ air between the molecules set by the Bernal limit, we can calculate an upper bound on the number of molecules.
\begin{equation}
  N \leq \frac{(\pi\times2\times 0.5^2)\times1.485\times 0.64}{6.55\times 10^{-19}} = 2.3\times10^{18} \text{ molecules},
\end{equation}
where the volume of the sapphire protrusion was subtracted.
After the experiment, we measured the mass of the \gls{dpph} in a weighing boat using a Shimadzu ATY224R placed on a mechanically isolated optical table.
This gave a lower bound reading of 1.0 mg, equivalent to $1.5\times 10^{18}$ molecules.

To calculate the number of spins in the paramagnetic phase, we use the sample temperature, which was held at 5.5 K by a proportional-integral-derivative (PID) controller.
Using Eq.~\ref{eq:paramag_fraction}, we determine that
\begin{equation}\label{eq:paramag_fraction_value}
  n_\mathrm{BB} = \frac{3e^{-\frac{17.6}{5.5}}}{1+3e^{-\frac{17.6}{5.5}}}\approx 0.109
\end{equation}
 of the spins will be in the triplet state.
The number of spins contributing to the \gls{epr} signal is calculated by multiplying this number by the polarization, given by\cite{reif2009fundamentals}
\begin{equation}
  P_B = \frac{1-e^{-\hbar\omega_s/k_B T}}{1+e^{-\hbar\omega_s/k_B T}}.
\end{equation}
For our resonator at $\omega_s/2\pi=4.557$~GHz and 5.5 K, this gives $P_B \approx 0.020$.
Multiplying $N$ by these last two factors gives the total number of spins contributing to the \gls{epr} signal,
\begin{equation}
  3.25\times 10^{15}<N_\mathrm{eff}<4.9\times 10^{15}.
\end{equation}

This is consistent with the simulated resonator-single spin coupling strength of 82~mHz and the measured ensemble coupling strength of 5.3~MHz.

\section{Free Induction Decay Simulation}
Because the \gls{fid} signal starts immediately after the pulse and because our short pulses have non-negligible slew rates, we simulate the system using the Quantum Toolbox in Python (QuTiP) with a time-dependent drive.
As we cannot simulate all $10^{15}$ spins, to simplify the simulation and reduce the computational cost, we first apply the Holstein-Primakoff approximation and truncate the bosonic modes at a few excitations.
Secondly, for \gls{fid} signals, we do not need to keep track of the phase of each spin in the inhomogeneously broadened ensemble because the dark states behave as a sink, in contrast to multi-pulse sequences where the ensemble is refocused.
The shape of the inhomogeneous broadening usually affects the shape of the \gls{fid}, but because $\kappa < \Gamma/2$ in all our experiments, we again treat the dark states as a Markovian bath.

This simulation is performed with the resonator, spin-ensemble and drive all on resonance and the interaction picture Hamiltonian
\begin{equation}
  H_\mathrm{int} = g_\mathrm{ens}(\hat{a}^\dag \hat{b} + \hat{a}\hat{b}^\dag) + i\sqrt{\kappa_\mathrm{ext}}\beta_{0}(t)(\hat{a}^\dag + \hat{a}),
\end{equation}
where the drive amplitude is time dependent.
In our case we truncate our Hilbert space at 15 levels for each bosonic mode, but any value will do as long as the drive is weak and short enough to keep the excitation low for the Holstein-Primakoff approximation.
The code can be found on Github\cite{couillardFidSimulation}

\bibliography{main,suppextra}

\begin{thebibliography}{89}%
\makeatletter
\providecommand \@ifxundefined [1]{%
 \@ifx{#1\undefined}
}%
\providecommand \@ifnum [1]{%
 \ifnum #1\expandafter \@firstoftwo
 \else \expandafter \@secondoftwo
 \fi
}%
\providecommand \@ifx [1]{%
 \ifx #1\expandafter \@firstoftwo
 \else \expandafter \@secondoftwo
 \fi
}%
\providecommand \natexlab [1]{#1}%
\providecommand \enquote  [1]{``#1''}%
\providecommand \bibnamefont  [1]{#1}%
\providecommand \bibfnamefont [1]{#1}%
\providecommand \citenamefont [1]{#1}%
\providecommand \href@noop [0]{\@secondoftwo}%
\providecommand \href [0]{\begingroup \@sanitize@url \@href}%
\providecommand \@href[1]{\@@startlink{#1}\@@href}%
\providecommand \@@href[1]{\endgroup#1\@@endlink}%
\providecommand \@sanitize@url [0]{\catcode `\\12\catcode `\$12\catcode `\&12\catcode `\#12\catcode `\^12\catcode `\_12\catcode `\%12\relax}%
\providecommand \@@startlink[1]{}%
\providecommand \@@endlink[0]{}%
\providecommand \url  [0]{\begingroup\@sanitize@url \@url }%
\providecommand \@url [1]{\endgroup\@href {#1}{\urlprefix }}%
\providecommand \urlprefix  [0]{URL }%
\providecommand \Eprint [0]{\href }%
\providecommand \doibase [0]{https://doi.org/}%
\providecommand \selectlanguage [0]{\@gobble}%
\providecommand \bibinfo  [0]{\@secondoftwo}%
\providecommand \bibfield  [0]{\@secondoftwo}%
\providecommand \translation [1]{[#1]}%
\providecommand \BibitemOpen [0]{}%
\providecommand \bibitemStop [0]{}%
\providecommand \bibitemNoStop [0]{.\EOS\space}%
\providecommand \EOS [0]{\spacefactor3000\relax}%
\providecommand \BibitemShut  [1]{\csname bibitem#1\endcsname}%
\let\auto@bib@innerbib\@empty
\bibitem [{\citenamefont {Lvovsky}\ \emph {et~al.}(2009)\citenamefont {Lvovsky}, \citenamefont {Sanders},\ and\ \citenamefont {Tittel}}]{lvovsky2009optical}%
  \BibitemOpen
  \bibfield  {author} {\bibinfo {author} {\bibfnamefont {A.~I.}\ \bibnamefont {Lvovsky}}, \bibinfo {author} {\bibfnamefont {B.~C.}\ \bibnamefont {Sanders}},\ and\ \bibinfo {author} {\bibfnamefont {W.}~\bibnamefont {Tittel}},\ }\bibfield  {title} {\bibinfo {title} {Optical quantum memory},\ }\href@noop {} {\bibfield  {journal} {\bibinfo  {journal} {Nature photonics}\ }\textbf {\bibinfo {volume} {3}},\ \bibinfo {pages} {706} (\bibinfo {year} {2009})}\BibitemShut {NoStop}%
\bibitem [{\citenamefont {Gorshkov}\ \emph {et~al.}(2007)\citenamefont {Gorshkov}, \citenamefont {Andr{\'e}}, \citenamefont {Lukin},\ and\ \citenamefont {S{\o}rensen}}]{gorshkov2007photon}%
  \BibitemOpen
  \bibfield  {author} {\bibinfo {author} {\bibfnamefont {A.~V.}\ \bibnamefont {Gorshkov}}, \bibinfo {author} {\bibfnamefont {A.}~\bibnamefont {Andr{\'e}}}, \bibinfo {author} {\bibfnamefont {M.~D.}\ \bibnamefont {Lukin}},\ and\ \bibinfo {author} {\bibfnamefont {A.~S.}\ \bibnamefont {S{\o}rensen}},\ }\bibfield  {title} {\bibinfo {title} {Photon storage in $\lambda$-type optically dense atomic media. ii. free-space model},\ }\href@noop {} {\bibfield  {journal} {\bibinfo  {journal} {Physical Review A—Atomic, Molecular, and Optical Physics}\ }\textbf {\bibinfo {volume} {76}},\ \bibinfo {pages} {033805} (\bibinfo {year} {2007})}\BibitemShut {NoStop}%
\bibitem [{\citenamefont {Afzelius}\ \emph {et~al.}(2009)\citenamefont {Afzelius}, \citenamefont {Simon}, \citenamefont {De~Riedmatten},\ and\ \citenamefont {Gisin}}]{afzelius2009multimode}%
  \BibitemOpen
  \bibfield  {author} {\bibinfo {author} {\bibfnamefont {M.}~\bibnamefont {Afzelius}}, \bibinfo {author} {\bibfnamefont {C.}~\bibnamefont {Simon}}, \bibinfo {author} {\bibfnamefont {H.}~\bibnamefont {De~Riedmatten}},\ and\ \bibinfo {author} {\bibfnamefont {N.}~\bibnamefont {Gisin}},\ }\bibfield  {title} {\bibinfo {title} {Multimode quantum memory based on atomic frequency combs},\ }\href@noop {} {\bibfield  {journal} {\bibinfo  {journal} {Physical Review A—Atomic, Molecular, and Optical Physics}\ }\textbf {\bibinfo {volume} {79}},\ \bibinfo {pages} {052329} (\bibinfo {year} {2009})}\BibitemShut {NoStop}%
\bibitem [{\citenamefont {Morton}\ \emph {et~al.}(2008)\citenamefont {Morton}, \citenamefont {Tyryshkin}, \citenamefont {Brown}, \citenamefont {Shankar}, \citenamefont {Lovett}, \citenamefont {Ardavan}, \citenamefont {Schenkel}, \citenamefont {Haller}, \citenamefont {Ager},\ and\ \citenamefont {Lyon}}]{morton2008solid}%
  \BibitemOpen
  \bibfield  {author} {\bibinfo {author} {\bibfnamefont {J.~J.}\ \bibnamefont {Morton}}, \bibinfo {author} {\bibfnamefont {A.~M.}\ \bibnamefont {Tyryshkin}}, \bibinfo {author} {\bibfnamefont {R.~M.}\ \bibnamefont {Brown}}, \bibinfo {author} {\bibfnamefont {S.}~\bibnamefont {Shankar}}, \bibinfo {author} {\bibfnamefont {B.~W.}\ \bibnamefont {Lovett}}, \bibinfo {author} {\bibfnamefont {A.}~\bibnamefont {Ardavan}}, \bibinfo {author} {\bibfnamefont {T.}~\bibnamefont {Schenkel}}, \bibinfo {author} {\bibfnamefont {E.~E.}\ \bibnamefont {Haller}}, \bibinfo {author} {\bibfnamefont {J.~W.}\ \bibnamefont {Ager}},\ and\ \bibinfo {author} {\bibfnamefont {S.}~\bibnamefont {Lyon}},\ }\bibfield  {title} {\bibinfo {title} {Solid-state quantum memory using the 31p nuclear spin},\ }\href@noop {} {\bibfield  {journal} {\bibinfo  {journal} {Nature}\ }\textbf {\bibinfo {volume} {455}},\ \bibinfo {pages} {1085} (\bibinfo {year} {2008})}\BibitemShut {NoStop}%
\bibitem [{\citenamefont {Wu}\ \emph {et~al.}(2010)\citenamefont {Wu}, \citenamefont {George}, \citenamefont {Wesenberg}, \citenamefont {M{\o}lmer}, \citenamefont {Schuster}, \citenamefont {Schoelkopf}, \citenamefont {Itoh}, \citenamefont {Ardavan}, \citenamefont {Morton},\ and\ \citenamefont {Briggs}}]{wu2010storage}%
  \BibitemOpen
  \bibfield  {author} {\bibinfo {author} {\bibfnamefont {H.}~\bibnamefont {Wu}}, \bibinfo {author} {\bibfnamefont {R.~E.}\ \bibnamefont {George}}, \bibinfo {author} {\bibfnamefont {J.~H.}\ \bibnamefont {Wesenberg}}, \bibinfo {author} {\bibfnamefont {K.}~\bibnamefont {M{\o}lmer}}, \bibinfo {author} {\bibfnamefont {D.~I.}\ \bibnamefont {Schuster}}, \bibinfo {author} {\bibfnamefont {R.~J.}\ \bibnamefont {Schoelkopf}}, \bibinfo {author} {\bibfnamefont {K.~M.}\ \bibnamefont {Itoh}}, \bibinfo {author} {\bibfnamefont {A.}~\bibnamefont {Ardavan}}, \bibinfo {author} {\bibfnamefont {J.~J.}\ \bibnamefont {Morton}},\ and\ \bibinfo {author} {\bibfnamefont {G.~A.~D.}\ \bibnamefont {Briggs}},\ }\bibfield  {title} {\bibinfo {title} {Storage of multiple coherent microwave excitations in an electron spin ensemble},\ }\href@noop {} {\bibfield  {journal} {\bibinfo  {journal} {Physical review letters}\ }\textbf {\bibinfo {volume} {105}},\ \bibinfo {pages} {140503} (\bibinfo {year} {2010})}\BibitemShut {NoStop}%
\bibitem [{\citenamefont {Kubo}\ \emph {et~al.}(2011)\citenamefont {Kubo}, \citenamefont {Grezes}, \citenamefont {Dewes}, \citenamefont {Umeda}, \citenamefont {Isoya}, \citenamefont {Sumiya}, \citenamefont {Morishita}, \citenamefont {Abe}, \citenamefont {Onoda}, \citenamefont {Ohshima}, \citenamefont {Jacques}, \citenamefont {Dr\'eau}, \citenamefont {Roch}, \citenamefont {Diniz}, \citenamefont {Auffeves}, \citenamefont {Vion}, \citenamefont {Esteve},\ and\ \citenamefont {Bertet}}]{kubo2011hybrid}%
  \BibitemOpen
  \bibfield  {author} {\bibinfo {author} {\bibfnamefont {Y.}~\bibnamefont {Kubo}}, \bibinfo {author} {\bibfnamefont {C.}~\bibnamefont {Grezes}}, \bibinfo {author} {\bibfnamefont {A.}~\bibnamefont {Dewes}}, \bibinfo {author} {\bibfnamefont {T.}~\bibnamefont {Umeda}}, \bibinfo {author} {\bibfnamefont {J.}~\bibnamefont {Isoya}}, \bibinfo {author} {\bibfnamefont {H.}~\bibnamefont {Sumiya}}, \bibinfo {author} {\bibfnamefont {N.}~\bibnamefont {Morishita}}, \bibinfo {author} {\bibfnamefont {H.}~\bibnamefont {Abe}}, \bibinfo {author} {\bibfnamefont {S.}~\bibnamefont {Onoda}}, \bibinfo {author} {\bibfnamefont {T.}~\bibnamefont {Ohshima}}, \bibinfo {author} {\bibfnamefont {V.}~\bibnamefont {Jacques}}, \bibinfo {author} {\bibfnamefont {A.}~\bibnamefont {Dr\'eau}}, \bibinfo {author} {\bibfnamefont {J.-F.}\ \bibnamefont {Roch}}, \bibinfo {author} {\bibfnamefont {I.}~\bibnamefont {Diniz}}, \bibinfo {author} {\bibfnamefont {A.}~\bibnamefont {Auffeves}}, \bibinfo {author} {\bibfnamefont {D.}~\bibnamefont {Vion}}, \bibinfo
  {author} {\bibfnamefont {D.}~\bibnamefont {Esteve}},\ and\ \bibinfo {author} {\bibfnamefont {P.}~\bibnamefont {Bertet}},\ }\bibfield  {title} {\bibinfo {title} {Hybrid quantum circuit with a superconducting qubit coupled to a spin ensemble},\ }\href {https://doi.org/10.1103/PhysRevLett.107.220501} {\bibfield  {journal} {\bibinfo  {journal} {Phys. Rev. Lett.}\ }\textbf {\bibinfo {volume} {107}},\ \bibinfo {pages} {220501} (\bibinfo {year} {2011})}\BibitemShut {NoStop}%
\bibitem [{\citenamefont {Zhu}\ \emph {et~al.}(2011)\citenamefont {Zhu}, \citenamefont {Saito}, \citenamefont {Kemp}, \citenamefont {Kakuyanagi}, \citenamefont {Karimoto}, \citenamefont {Nakano}, \citenamefont {Munro}, \citenamefont {Tokura}, \citenamefont {Everitt}, \citenamefont {Nemoto}, \citenamefont {Kasu}, \citenamefont {Mizuochi},\ and\ \citenamefont {Semba}}]{zhu2011coherent}%
  \BibitemOpen
  \bibfield  {author} {\bibinfo {author} {\bibfnamefont {X.}~\bibnamefont {Zhu}}, \bibinfo {author} {\bibfnamefont {S.}~\bibnamefont {Saito}}, \bibinfo {author} {\bibfnamefont {A.}~\bibnamefont {Kemp}}, \bibinfo {author} {\bibfnamefont {K.}~\bibnamefont {Kakuyanagi}}, \bibinfo {author} {\bibfnamefont {S.-i.}\ \bibnamefont {Karimoto}}, \bibinfo {author} {\bibfnamefont {H.}~\bibnamefont {Nakano}}, \bibinfo {author} {\bibfnamefont {W.~J.}\ \bibnamefont {Munro}}, \bibinfo {author} {\bibfnamefont {Y.}~\bibnamefont {Tokura}}, \bibinfo {author} {\bibfnamefont {M.~S.}\ \bibnamefont {Everitt}}, \bibinfo {author} {\bibfnamefont {K.}~\bibnamefont {Nemoto}}, \bibinfo {author} {\bibfnamefont {M.}~\bibnamefont {Kasu}}, \bibinfo {author} {\bibfnamefont {N.}~\bibnamefont {Mizuochi}},\ and\ \bibinfo {author} {\bibfnamefont {K.}~\bibnamefont {Semba}},\ }\bibfield  {title} {\bibinfo {title} {Coherent coupling of a superconducting flux qubit to an electron spin ensemble in diamond},\ }\href {https://doi.org/10.1038/nature10462}
  {\bibfield  {journal} {\bibinfo  {journal} {Nature}\ }\textbf {\bibinfo {volume} {478}},\ \bibinfo {pages} {221} (\bibinfo {year} {2011})}\BibitemShut {NoStop}%
\bibitem [{\citenamefont {Julsgaard}\ \emph {et~al.}(2013)\citenamefont {Julsgaard}, \citenamefont {Grezes}, \citenamefont {Bertet},\ and\ \citenamefont {M{\o}lmer}}]{julsgaard2013quantum}%
  \BibitemOpen
  \bibfield  {author} {\bibinfo {author} {\bibfnamefont {B.}~\bibnamefont {Julsgaard}}, \bibinfo {author} {\bibfnamefont {C.}~\bibnamefont {Grezes}}, \bibinfo {author} {\bibfnamefont {P.}~\bibnamefont {Bertet}},\ and\ \bibinfo {author} {\bibfnamefont {K.}~\bibnamefont {M{\o}lmer}},\ }\bibfield  {title} {\bibinfo {title} {Quantum memory for microwave photons in an inhomogeneously broadened spin ensemble},\ }\href@noop {} {\bibfield  {journal} {\bibinfo  {journal} {Physical review letters}\ }\textbf {\bibinfo {volume} {110}},\ \bibinfo {pages} {250503} (\bibinfo {year} {2013})}\BibitemShut {NoStop}%
\bibitem [{\citenamefont {Afzelius}\ \emph {et~al.}(2013)\citenamefont {Afzelius}, \citenamefont {Sangouard}, \citenamefont {Johansson}, \citenamefont {Staudt},\ and\ \citenamefont {Wilson}}]{afzelius2013proposal}%
  \BibitemOpen
  \bibfield  {author} {\bibinfo {author} {\bibfnamefont {M.}~\bibnamefont {Afzelius}}, \bibinfo {author} {\bibfnamefont {N.}~\bibnamefont {Sangouard}}, \bibinfo {author} {\bibfnamefont {G.}~\bibnamefont {Johansson}}, \bibinfo {author} {\bibfnamefont {M.}~\bibnamefont {Staudt}},\ and\ \bibinfo {author} {\bibfnamefont {C.}~\bibnamefont {Wilson}},\ }\bibfield  {title} {\bibinfo {title} {Proposal for a coherent quantum memory for propagating microwave photons},\ }\href@noop {} {\bibfield  {journal} {\bibinfo  {journal} {New Journal of Physics}\ }\textbf {\bibinfo {volume} {15}},\ \bibinfo {pages} {065008} (\bibinfo {year} {2013})}\BibitemShut {NoStop}%
\bibitem [{\citenamefont {Grezes}\ \emph {et~al.}(2014)\citenamefont {Grezes}, \citenamefont {Julsgaard}, \citenamefont {Kubo}, \citenamefont {Stern}, \citenamefont {Umeda}, \citenamefont {Isoya}, \citenamefont {Sumiya}, \citenamefont {Abe}, \citenamefont {Onoda}, \citenamefont {Ohshima} \emph {et~al.}}]{grezes2014multimode}%
  \BibitemOpen
  \bibfield  {author} {\bibinfo {author} {\bibfnamefont {C.}~\bibnamefont {Grezes}}, \bibinfo {author} {\bibfnamefont {B.}~\bibnamefont {Julsgaard}}, \bibinfo {author} {\bibfnamefont {Y.}~\bibnamefont {Kubo}}, \bibinfo {author} {\bibfnamefont {M.}~\bibnamefont {Stern}}, \bibinfo {author} {\bibfnamefont {T.}~\bibnamefont {Umeda}}, \bibinfo {author} {\bibfnamefont {J.}~\bibnamefont {Isoya}}, \bibinfo {author} {\bibfnamefont {H.}~\bibnamefont {Sumiya}}, \bibinfo {author} {\bibfnamefont {H.}~\bibnamefont {Abe}}, \bibinfo {author} {\bibfnamefont {S.}~\bibnamefont {Onoda}}, \bibinfo {author} {\bibfnamefont {T.}~\bibnamefont {Ohshima}}, \emph {et~al.},\ }\bibfield  {title} {\bibinfo {title} {Multimode storage and retrieval of microwave fields in a spin ensemble},\ }\href@noop {} {\bibfield  {journal} {\bibinfo  {journal} {Physical Review X}\ }\textbf {\bibinfo {volume} {4}},\ \bibinfo {pages} {021049} (\bibinfo {year} {2014})}\BibitemShut {NoStop}%
\bibitem [{\citenamefont {Ranjan}\ \emph {et~al.}(2020{\natexlab{a}})\citenamefont {Ranjan}, \citenamefont {O’sullivan}, \citenamefont {Albertinale}, \citenamefont {Albanese}, \citenamefont {Chaneli{\`e}re}, \citenamefont {Schenkel}, \citenamefont {Vion}, \citenamefont {Esteve}, \citenamefont {Flurin}, \citenamefont {Morton} \emph {et~al.}}]{ranjan2020multimode}%
  \BibitemOpen
  \bibfield  {author} {\bibinfo {author} {\bibfnamefont {V.}~\bibnamefont {Ranjan}}, \bibinfo {author} {\bibfnamefont {J.}~\bibnamefont {O’sullivan}}, \bibinfo {author} {\bibfnamefont {E.}~\bibnamefont {Albertinale}}, \bibinfo {author} {\bibfnamefont {B.}~\bibnamefont {Albanese}}, \bibinfo {author} {\bibfnamefont {T.}~\bibnamefont {Chaneli{\`e}re}}, \bibinfo {author} {\bibfnamefont {T.}~\bibnamefont {Schenkel}}, \bibinfo {author} {\bibfnamefont {D.}~\bibnamefont {Vion}}, \bibinfo {author} {\bibfnamefont {D.}~\bibnamefont {Esteve}}, \bibinfo {author} {\bibfnamefont {E.}~\bibnamefont {Flurin}}, \bibinfo {author} {\bibfnamefont {J.}~\bibnamefont {Morton}}, \emph {et~al.},\ }\bibfield  {title} {\bibinfo {title} {Multimode storage of quantum microwave fields in electron spins over 100 ms},\ }\href@noop {} {\bibfield  {journal} {\bibinfo  {journal} {Physical Review Letters}\ }\textbf {\bibinfo {volume} {125}},\ \bibinfo {pages} {210505} (\bibinfo {year} {2020}{\natexlab{a}})}\BibitemShut {NoStop}%
\bibitem [{\citenamefont {Greggio}\ \emph {et~al.}(2025)\citenamefont {Greggio}, \citenamefont {Lorriaux}, \citenamefont {Petrescu}, \citenamefont {Mirrahimi},\ and\ \citenamefont {Bienfait}}]{greggio2025optimal}%
  \BibitemOpen
  \bibfield  {author} {\bibinfo {author} {\bibfnamefont {L.}~\bibnamefont {Greggio}}, \bibinfo {author} {\bibfnamefont {T.}~\bibnamefont {Lorriaux}}, \bibinfo {author} {\bibfnamefont {A.}~\bibnamefont {Petrescu}}, \bibinfo {author} {\bibfnamefont {M.}~\bibnamefont {Mirrahimi}},\ and\ \bibinfo {author} {\bibfnamefont {A.}~\bibnamefont {Bienfait}},\ }\bibfield  {title} {\bibinfo {title} {Optimal absorption and emission of itinerant fields into a spin ensemble memory},\ }\href@noop {} {\bibfield  {journal} {\bibinfo  {journal} {arXiv preprint arXiv:2506.06107}\ } (\bibinfo {year} {2025})}\BibitemShut {NoStop}%
\bibitem [{\citenamefont {Lambert}\ \emph {et~al.}(2020)\citenamefont {Lambert}, \citenamefont {Rueda}, \citenamefont {Sedlmeir},\ and\ \citenamefont {Schwefel}}]{lambert2020coherent}%
  \BibitemOpen
  \bibfield  {author} {\bibinfo {author} {\bibfnamefont {N.~J.}\ \bibnamefont {Lambert}}, \bibinfo {author} {\bibfnamefont {A.}~\bibnamefont {Rueda}}, \bibinfo {author} {\bibfnamefont {F.}~\bibnamefont {Sedlmeir}},\ and\ \bibinfo {author} {\bibfnamefont {H.~G.~L.}\ \bibnamefont {Schwefel}},\ }\bibfield  {title} {\bibinfo {title} {Coherent conversion between microwave and optical photons---an overview of physical implementations},\ }\href {https://doi.org/10.1002/qute.201900077} {\bibfield  {journal} {\bibinfo  {journal} {Advanced Quantum Technologies}\ }\textbf {\bibinfo {volume} {3}},\ \bibinfo {pages} {1900077} (\bibinfo {year} {2020})}\BibitemShut {NoStop}%
\bibitem [{\citenamefont {Hafezi}\ \emph {et~al.}(2012)\citenamefont {Hafezi}, \citenamefont {Kim}, \citenamefont {Rolston}, \citenamefont {Orozco}, \citenamefont {Lev},\ and\ \citenamefont {Taylor}}]{hafezi2012atomic}%
  \BibitemOpen
  \bibfield  {author} {\bibinfo {author} {\bibfnamefont {M.}~\bibnamefont {Hafezi}}, \bibinfo {author} {\bibfnamefont {Z.}~\bibnamefont {Kim}}, \bibinfo {author} {\bibfnamefont {S.~L.}\ \bibnamefont {Rolston}}, \bibinfo {author} {\bibfnamefont {L.~A.}\ \bibnamefont {Orozco}}, \bibinfo {author} {\bibfnamefont {B.}~\bibnamefont {Lev}},\ and\ \bibinfo {author} {\bibfnamefont {J.~M.}\ \bibnamefont {Taylor}},\ }\bibfield  {title} {\bibinfo {title} {Atomic interface between microwave and optical photons},\ }\href@noop {} {\bibfield  {journal} {\bibinfo  {journal} {Physical Review A—Atomic, Molecular, and Optical Physics}\ }\textbf {\bibinfo {volume} {85}},\ \bibinfo {pages} {020302} (\bibinfo {year} {2012})}\BibitemShut {NoStop}%
\bibitem [{\citenamefont {O'Brien}\ \emph {et~al.}(2014)\citenamefont {O'Brien}, \citenamefont {Lauk}, \citenamefont {Blum}, \citenamefont {Morigi},\ and\ \citenamefont {Fleischhauer}}]{o2014interfacing}%
  \BibitemOpen
  \bibfield  {author} {\bibinfo {author} {\bibfnamefont {C.}~\bibnamefont {O'Brien}}, \bibinfo {author} {\bibfnamefont {N.}~\bibnamefont {Lauk}}, \bibinfo {author} {\bibfnamefont {S.}~\bibnamefont {Blum}}, \bibinfo {author} {\bibfnamefont {G.}~\bibnamefont {Morigi}},\ and\ \bibinfo {author} {\bibfnamefont {M.}~\bibnamefont {Fleischhauer}},\ }\bibfield  {title} {\bibinfo {title} {Interfacing superconducting qubits and telecom photons via a rare-earth doped crystal},\ }\href@noop {} {\bibfield  {journal} {\bibinfo  {journal} {arXiv preprint arXiv:1402.5405}\ } (\bibinfo {year} {2014})}\BibitemShut {NoStop}%
\bibitem [{\citenamefont {Williamson}\ \emph {et~al.}(2014)\citenamefont {Williamson}, \citenamefont {Chen},\ and\ \citenamefont {Longdell}}]{williamson2014magneto}%
  \BibitemOpen
  \bibfield  {author} {\bibinfo {author} {\bibfnamefont {L.~A.}\ \bibnamefont {Williamson}}, \bibinfo {author} {\bibfnamefont {Y.-H.}\ \bibnamefont {Chen}},\ and\ \bibinfo {author} {\bibfnamefont {J.~J.}\ \bibnamefont {Longdell}},\ }\bibfield  {title} {\bibinfo {title} {Magneto-optic modulator with unit quantum efficiency},\ }\href@noop {} {\bibfield  {journal} {\bibinfo  {journal} {Physical review letters}\ }\textbf {\bibinfo {volume} {113}},\ \bibinfo {pages} {203601} (\bibinfo {year} {2014})}\BibitemShut {NoStop}%
\bibitem [{\citenamefont {Fernandez-Gonzalvo}\ \emph {et~al.}(2015)\citenamefont {Fernandez-Gonzalvo}, \citenamefont {Chen}, \citenamefont {Yin}, \citenamefont {Rogge},\ and\ \citenamefont {Longdell}}]{fernandezgonzalvo2015coherent}%
  \BibitemOpen
  \bibfield  {author} {\bibinfo {author} {\bibfnamefont {X.}~\bibnamefont {Fernandez-Gonzalvo}}, \bibinfo {author} {\bibfnamefont {Y.-H.}\ \bibnamefont {Chen}}, \bibinfo {author} {\bibfnamefont {C.}~\bibnamefont {Yin}}, \bibinfo {author} {\bibfnamefont {S.}~\bibnamefont {Rogge}},\ and\ \bibinfo {author} {\bibfnamefont {J.~J.}\ \bibnamefont {Longdell}},\ }\bibfield  {title} {\bibinfo {title} {Coherent frequency up-conversion of microwaves to the optical telecommunications band in an {Er:YSO} crystal},\ }\href {https://doi.org/10.1103/PhysRevA.92.062313} {\bibfield  {journal} {\bibinfo  {journal} {Physical Review A}\ }\textbf {\bibinfo {volume} {92}},\ \bibinfo {pages} {062313} (\bibinfo {year} {2015})}\BibitemShut {NoStop}%
\bibitem [{\citenamefont {Hisatomi}\ \emph {et~al.}(2016)\citenamefont {Hisatomi}, \citenamefont {Osada}, \citenamefont {Tabuchi}, \citenamefont {Ishikawa}, \citenamefont {Noguchi}, \citenamefont {Yamazaki}, \citenamefont {Usami},\ and\ \citenamefont {Nakamura}}]{hisatomi2016bidirectional}%
  \BibitemOpen
  \bibfield  {author} {\bibinfo {author} {\bibfnamefont {R.}~\bibnamefont {Hisatomi}}, \bibinfo {author} {\bibfnamefont {A.}~\bibnamefont {Osada}}, \bibinfo {author} {\bibfnamefont {Y.}~\bibnamefont {Tabuchi}}, \bibinfo {author} {\bibfnamefont {T.}~\bibnamefont {Ishikawa}}, \bibinfo {author} {\bibfnamefont {A.}~\bibnamefont {Noguchi}}, \bibinfo {author} {\bibfnamefont {R.}~\bibnamefont {Yamazaki}}, \bibinfo {author} {\bibfnamefont {K.}~\bibnamefont {Usami}},\ and\ \bibinfo {author} {\bibfnamefont {Y.}~\bibnamefont {Nakamura}},\ }\bibfield  {title} {\bibinfo {title} {Bidirectional conversion between microwave and light via ferromagnetic magnons},\ }\href {https://doi.org/10.1103/PhysRevB.93.174427} {\bibfield  {journal} {\bibinfo  {journal} {Physical Review B}\ }\textbf {\bibinfo {volume} {93}},\ \bibinfo {pages} {174427} (\bibinfo {year} {2016})}\BibitemShut {NoStop}%
\bibitem [{\citenamefont {Fernandez-Gonzalvo}\ \emph {et~al.}(2019)\citenamefont {Fernandez-Gonzalvo}, \citenamefont {Horvath}, \citenamefont {Chen},\ and\ \citenamefont {Longdell}}]{fernandezgonzalvo2019cavity}%
  \BibitemOpen
  \bibfield  {author} {\bibinfo {author} {\bibfnamefont {X.}~\bibnamefont {Fernandez-Gonzalvo}}, \bibinfo {author} {\bibfnamefont {S.~P.}\ \bibnamefont {Horvath}}, \bibinfo {author} {\bibfnamefont {Y.-H.}\ \bibnamefont {Chen}},\ and\ \bibinfo {author} {\bibfnamefont {J.~J.}\ \bibnamefont {Longdell}},\ }\bibfield  {title} {\bibinfo {title} {Cavity-enhanced raman heterodyne spectroscopy in {Er$^{3+}$:Y$_2$SiO$_5$} for microwave to optical signal conversion},\ }\href {https://doi.org/10.1103/PhysRevA.100.033807} {\bibfield  {journal} {\bibinfo  {journal} {Physical Review A}\ }\textbf {\bibinfo {volume} {100}},\ \bibinfo {pages} {033807} (\bibinfo {year} {2019})}\BibitemShut {NoStop}%
\bibitem [{\citenamefont {Barnett}\ and\ \citenamefont {Longdell}(2020)}]{barnett2020theory}%
  \BibitemOpen
  \bibfield  {author} {\bibinfo {author} {\bibfnamefont {P.~S.}\ \bibnamefont {Barnett}}\ and\ \bibinfo {author} {\bibfnamefont {J.~J.}\ \bibnamefont {Longdell}},\ }\bibfield  {title} {\bibinfo {title} {Theory of microwave-optical conversion using rare-earth-ion dopants},\ }\href {https://doi.org/10.1103/PhysRevA.102.063718} {\bibfield  {journal} {\bibinfo  {journal} {Physical Review A}\ }\textbf {\bibinfo {volume} {102}},\ \bibinfo {pages} {063718} (\bibinfo {year} {2020})}\BibitemShut {NoStop}%
\bibitem [{\citenamefont {Bartholomew}\ \emph {et~al.}(2020)\citenamefont {Bartholomew}, \citenamefont {Rochman}, \citenamefont {Xie}, \citenamefont {Kindem}, \citenamefont {Ruskuc}, \citenamefont {Craiciu}, \citenamefont {Lei},\ and\ \citenamefont {Faraon}}]{bartholomew2020chip}%
  \BibitemOpen
  \bibfield  {author} {\bibinfo {author} {\bibfnamefont {J.~G.}\ \bibnamefont {Bartholomew}}, \bibinfo {author} {\bibfnamefont {J.}~\bibnamefont {Rochman}}, \bibinfo {author} {\bibfnamefont {T.}~\bibnamefont {Xie}}, \bibinfo {author} {\bibfnamefont {J.~M.}\ \bibnamefont {Kindem}}, \bibinfo {author} {\bibfnamefont {A.}~\bibnamefont {Ruskuc}}, \bibinfo {author} {\bibfnamefont {I.}~\bibnamefont {Craiciu}}, \bibinfo {author} {\bibfnamefont {M.}~\bibnamefont {Lei}},\ and\ \bibinfo {author} {\bibfnamefont {A.}~\bibnamefont {Faraon}},\ }\bibfield  {title} {\bibinfo {title} {On-chip coherent microwave-to-optical transduction mediated by ytterbium in {YVO$_4$}},\ }\href {https://doi.org/10.1038/s41467-020-16996-x} {\bibfield  {journal} {\bibinfo  {journal} {Nature Communications}\ }\textbf {\bibinfo {volume} {11}},\ \bibinfo {pages} {3266} (\bibinfo {year} {2020})}\BibitemShut {NoStop}%
\bibitem [{\citenamefont {Li}\ \emph {et~al.}(2023)\citenamefont {Li}, \citenamefont {Du}, \citenamefont {Wilson},\ and\ \citenamefont {Bajcsy}}]{li2023fiber}%
  \BibitemOpen
  \bibfield  {author} {\bibinfo {author} {\bibfnamefont {W.}~\bibnamefont {Li}}, \bibinfo {author} {\bibfnamefont {J.}~\bibnamefont {Du}}, \bibinfo {author} {\bibfnamefont {C.~M.}\ \bibnamefont {Wilson}},\ and\ \bibinfo {author} {\bibfnamefont {M.}~\bibnamefont {Bajcsy}},\ }\bibfield  {title} {\bibinfo {title} {Fiber-integrated microwave-to-optical quantum transducer},\ }\href {https://doi.org/10.1103/PhysRevApplied.20.044031} {\bibfield  {journal} {\bibinfo  {journal} {Physical Review Applied}\ }\textbf {\bibinfo {volume} {20}},\ \bibinfo {pages} {044031} (\bibinfo {year} {2023})}\BibitemShut {NoStop}%
\bibitem [{\citenamefont {Rochman}\ \emph {et~al.}(2023)\citenamefont {Rochman}, \citenamefont {Xie}, \citenamefont {Bartholomew}, \citenamefont {Schwab},\ and\ \citenamefont {Faraon}}]{rochman2023microwave}%
  \BibitemOpen
  \bibfield  {author} {\bibinfo {author} {\bibfnamefont {J.}~\bibnamefont {Rochman}}, \bibinfo {author} {\bibfnamefont {T.}~\bibnamefont {Xie}}, \bibinfo {author} {\bibfnamefont {J.~G.}\ \bibnamefont {Bartholomew}}, \bibinfo {author} {\bibfnamefont {K.~C.}\ \bibnamefont {Schwab}},\ and\ \bibinfo {author} {\bibfnamefont {A.}~\bibnamefont {Faraon}},\ }\bibfield  {title} {\bibinfo {title} {Microwave-to-optical transduction with erbium ions coupled to planar photonic and superconducting resonators},\ }\href {https://doi.org/10.1038/s41467-023-36799-0} {\bibfield  {journal} {\bibinfo  {journal} {Nature Communications}\ }\textbf {\bibinfo {volume} {14}},\ \bibinfo {pages} {1153} (\bibinfo {year} {2023})}\BibitemShut {NoStop}%
\bibitem [{\citenamefont {Xie}\ \emph {et~al.}(2025)\citenamefont {Xie}, \citenamefont {Fukumori}, \citenamefont {Li},\ and\ \citenamefont {Faraon}}]{xie2025scalable}%
  \BibitemOpen
  \bibfield  {author} {\bibinfo {author} {\bibfnamefont {T.}~\bibnamefont {Xie}}, \bibinfo {author} {\bibfnamefont {R.}~\bibnamefont {Fukumori}}, \bibinfo {author} {\bibfnamefont {J.}~\bibnamefont {Li}},\ and\ \bibinfo {author} {\bibfnamefont {A.}~\bibnamefont {Faraon}},\ }\bibfield  {title} {\bibinfo {title} {Scalable microwave-to-optical transducers at the single-photon level with spins},\ }\href {https://doi.org/10.1038/s41567-025-02884-y} {\bibfield  {journal} {\bibinfo  {journal} {Nature Physics}\ }\textbf {\bibinfo {volume} {21}},\ \bibinfo {pages} {931} (\bibinfo {year} {2025})}\BibitemShut {NoStop}%
\bibitem [{\citenamefont {Sherman}\ \emph {et~al.}(2022)\citenamefont {Sherman}, \citenamefont {Zgadzai}, \citenamefont {Koren}, \citenamefont {Peretz}, \citenamefont {Laster},\ and\ \citenamefont {Blank}}]{sherman2022diamond}%
  \BibitemOpen
  \bibfield  {author} {\bibinfo {author} {\bibfnamefont {A.}~\bibnamefont {Sherman}}, \bibinfo {author} {\bibfnamefont {O.}~\bibnamefont {Zgadzai}}, \bibinfo {author} {\bibfnamefont {B.}~\bibnamefont {Koren}}, \bibinfo {author} {\bibfnamefont {I.}~\bibnamefont {Peretz}}, \bibinfo {author} {\bibfnamefont {E.}~\bibnamefont {Laster}},\ and\ \bibinfo {author} {\bibfnamefont {A.}~\bibnamefont {Blank}},\ }\bibfield  {title} {\bibinfo {title} {Diamond-based microwave quantum amplifier},\ }\href {https://doi.org/10.1126/sciadv.ade6527} {\bibfield  {journal} {\bibinfo  {journal} {Science Advances}\ }\textbf {\bibinfo {volume} {8}},\ \bibinfo {pages} {eade6527} (\bibinfo {year} {2022})}\BibitemShut {NoStop}%
\bibitem [{\citenamefont {Day}\ \emph {et~al.}(2024)\citenamefont {Day}, \citenamefont {Isarov}, \citenamefont {Pappas}, \citenamefont {Johnson}, \citenamefont {Abe}, \citenamefont {Ohshima}, \citenamefont {McCamey}, \citenamefont {Laucht},\ and\ \citenamefont {Pla}}]{day2024room}%
  \BibitemOpen
  \bibfield  {author} {\bibinfo {author} {\bibfnamefont {T.}~\bibnamefont {Day}}, \bibinfo {author} {\bibfnamefont {M.}~\bibnamefont {Isarov}}, \bibinfo {author} {\bibfnamefont {W.~J.}\ \bibnamefont {Pappas}}, \bibinfo {author} {\bibfnamefont {B.~C.}\ \bibnamefont {Johnson}}, \bibinfo {author} {\bibfnamefont {H.}~\bibnamefont {Abe}}, \bibinfo {author} {\bibfnamefont {T.}~\bibnamefont {Ohshima}}, \bibinfo {author} {\bibfnamefont {D.~R.}\ \bibnamefont {McCamey}}, \bibinfo {author} {\bibfnamefont {A.}~\bibnamefont {Laucht}},\ and\ \bibinfo {author} {\bibfnamefont {J.~J.}\ \bibnamefont {Pla}},\ }\bibfield  {title} {\bibinfo {title} {Room-temperature solid-state maser amplifier},\ }\href@noop {} {\bibfield  {journal} {\bibinfo  {journal} {Physical Review X}\ }\textbf {\bibinfo {volume} {14}},\ \bibinfo {pages} {041066} (\bibinfo {year} {2024})}\BibitemShut {NoStop}%
\bibitem [{\citenamefont {Ohta}\ \emph {et~al.}(2025)\citenamefont {Ohta}, \citenamefont {Lee}, \citenamefont {Sietses}, \citenamefont {Kostylev}, \citenamefont {Ball}, \citenamefont {Moroshkin}, \citenamefont {Hamamoto}, \citenamefont {Kobayashi}, \citenamefont {Onoda}, \citenamefont {Ohshima} \emph {et~al.}}]{ohta2025near}%
  \BibitemOpen
  \bibfield  {author} {\bibinfo {author} {\bibfnamefont {M.}~\bibnamefont {Ohta}}, \bibinfo {author} {\bibfnamefont {C.-P.}\ \bibnamefont {Lee}}, \bibinfo {author} {\bibfnamefont {V.~P.}\ \bibnamefont {Sietses}}, \bibinfo {author} {\bibfnamefont {I.}~\bibnamefont {Kostylev}}, \bibinfo {author} {\bibfnamefont {J.~R.}\ \bibnamefont {Ball}}, \bibinfo {author} {\bibfnamefont {P.}~\bibnamefont {Moroshkin}}, \bibinfo {author} {\bibfnamefont {T.}~\bibnamefont {Hamamoto}}, \bibinfo {author} {\bibfnamefont {Y.}~\bibnamefont {Kobayashi}}, \bibinfo {author} {\bibfnamefont {S.}~\bibnamefont {Onoda}}, \bibinfo {author} {\bibfnamefont {T.}~\bibnamefont {Ohshima}}, \emph {et~al.},\ }\bibfield  {title} {\bibinfo {title} {A near-quantum-limited diamond maser amplifier operating at millikelvin temperatures},\ }\href@noop {} {\bibfield  {journal} {\bibinfo  {journal} {arXiv preprint arXiv:2505.05705}\ } (\bibinfo {year} {2025})}\BibitemShut {NoStop}%
\bibitem [{\citenamefont {Degen}\ \emph {et~al.}(2017)\citenamefont {Degen}, \citenamefont {Reinhard},\ and\ \citenamefont {Cappellaro}}]{degen2017quantum}%
  \BibitemOpen
  \bibfield  {author} {\bibinfo {author} {\bibfnamefont {C.~L.}\ \bibnamefont {Degen}}, \bibinfo {author} {\bibfnamefont {F.}~\bibnamefont {Reinhard}},\ and\ \bibinfo {author} {\bibfnamefont {P.}~\bibnamefont {Cappellaro}},\ }\bibfield  {title} {\bibinfo {title} {Quantum sensing},\ }\href {https://doi.org/10.1103/RevModPhys.89.035002} {\bibfield  {journal} {\bibinfo  {journal} {Reviews of Modern Physics}\ }\textbf {\bibinfo {volume} {89}},\ \bibinfo {pages} {035002} (\bibinfo {year} {2017})}\BibitemShut {NoStop}%
\bibitem [{\citenamefont {Barry}\ \emph {et~al.}(2020)\citenamefont {Barry}, \citenamefont {Schloss}, \citenamefont {Bauch}, \citenamefont {Turner}, \citenamefont {Hart}, \citenamefont {Pham},\ and\ \citenamefont {Walsworth}}]{barry2020sensitivity}%
  \BibitemOpen
  \bibfield  {author} {\bibinfo {author} {\bibfnamefont {J.~F.}\ \bibnamefont {Barry}}, \bibinfo {author} {\bibfnamefont {J.~M.}\ \bibnamefont {Schloss}}, \bibinfo {author} {\bibfnamefont {E.}~\bibnamefont {Bauch}}, \bibinfo {author} {\bibfnamefont {M.~J.}\ \bibnamefont {Turner}}, \bibinfo {author} {\bibfnamefont {C.~A.}\ \bibnamefont {Hart}}, \bibinfo {author} {\bibfnamefont {L.~M.}\ \bibnamefont {Pham}},\ and\ \bibinfo {author} {\bibfnamefont {R.~L.}\ \bibnamefont {Walsworth}},\ }\bibfield  {title} {\bibinfo {title} {Sensitivity optimization for {NV}-diamond magnetometry},\ }\href {https://doi.org/10.1103/RevModPhys.92.015004} {\bibfield  {journal} {\bibinfo  {journal} {Reviews of Modern Physics}\ }\textbf {\bibinfo {volume} {92}},\ \bibinfo {pages} {015004} (\bibinfo {year} {2020})}\BibitemShut {NoStop}%
\bibitem [{\citenamefont {Michl}\ \emph {et~al.}(2019)\citenamefont {Michl}, \citenamefont {Steiner}, \citenamefont {Denisenko}, \citenamefont {B{\"u}lau}, \citenamefont {Zimmermann}, \citenamefont {Nakamura}, \citenamefont {Sumiya}, \citenamefont {Onoda}, \citenamefont {Neumann}, \citenamefont {Isoya} \emph {et~al.}}]{michl2019robust}%
  \BibitemOpen
  \bibfield  {author} {\bibinfo {author} {\bibfnamefont {J.}~\bibnamefont {Michl}}, \bibinfo {author} {\bibfnamefont {J.}~\bibnamefont {Steiner}}, \bibinfo {author} {\bibfnamefont {A.}~\bibnamefont {Denisenko}}, \bibinfo {author} {\bibfnamefont {A.}~\bibnamefont {B{\"u}lau}}, \bibinfo {author} {\bibfnamefont {A.}~\bibnamefont {Zimmermann}}, \bibinfo {author} {\bibfnamefont {K.}~\bibnamefont {Nakamura}}, \bibinfo {author} {\bibfnamefont {H.}~\bibnamefont {Sumiya}}, \bibinfo {author} {\bibfnamefont {S.}~\bibnamefont {Onoda}}, \bibinfo {author} {\bibfnamefont {P.}~\bibnamefont {Neumann}}, \bibinfo {author} {\bibfnamefont {J.}~\bibnamefont {Isoya}}, \emph {et~al.},\ }\bibfield  {title} {\bibinfo {title} {Robust and accurate electric field sensing with solid state spin ensembles},\ }\href@noop {} {\bibfield  {journal} {\bibinfo  {journal} {Nano letters}\ }\textbf {\bibinfo {volume} {19}},\ \bibinfo {pages} {4904} (\bibinfo {year} {2019})}\BibitemShut {NoStop}%
\bibitem [{\citenamefont {Fan}\ \emph {et~al.}(2015)\citenamefont {Fan}, \citenamefont {Kumar}, \citenamefont {Sedlacek}, \citenamefont {K{\\\"u}bler}, \citenamefont {Karimkashi},\ and\ \citenamefont {Shaffer}}]{fan2015atom}%
  \BibitemOpen
  \bibfield  {author} {\bibinfo {author} {\bibfnamefont {H.}~\bibnamefont {Fan}}, \bibinfo {author} {\bibfnamefont {S.}~\bibnamefont {Kumar}}, \bibinfo {author} {\bibfnamefont {J.}~\bibnamefont {Sedlacek}}, \bibinfo {author} {\bibfnamefont {H.}~\bibnamefont {K{\\\"u}bler}}, \bibinfo {author} {\bibfnamefont {S.}~\bibnamefont {Karimkashi}},\ and\ \bibinfo {author} {\bibfnamefont {J.~P.}\ \bibnamefont {Shaffer}},\ }\bibfield  {title} {\bibinfo {title} {Atom based rf electric field sensing},\ }\href@noop {} {\bibfield  {journal} {\bibinfo  {journal} {Journal of Physics B: Atomic, Molecular and Optical Physics}\ }\textbf {\bibinfo {volume} {48}},\ \bibinfo {pages} {202001} (\bibinfo {year} {2015})}\BibitemShut {NoStop}%
\bibitem [{\citenamefont {Hong}\ \emph {et~al.}(2013)\citenamefont {Hong}, \citenamefont {Grinolds}, \citenamefont {Pham}, \citenamefont {Le~Sage}, \citenamefont {Luan}, \citenamefont {Walsworth},\ and\ \citenamefont {Yacoby}}]{hong2013nanoscale}%
  \BibitemOpen
  \bibfield  {author} {\bibinfo {author} {\bibfnamefont {S.}~\bibnamefont {Hong}}, \bibinfo {author} {\bibfnamefont {M.~S.}\ \bibnamefont {Grinolds}}, \bibinfo {author} {\bibfnamefont {L.~M.}\ \bibnamefont {Pham}}, \bibinfo {author} {\bibfnamefont {D.}~\bibnamefont {Le~Sage}}, \bibinfo {author} {\bibfnamefont {L.}~\bibnamefont {Luan}}, \bibinfo {author} {\bibfnamefont {R.~L.}\ \bibnamefont {Walsworth}},\ and\ \bibinfo {author} {\bibfnamefont {A.}~\bibnamefont {Yacoby}},\ }\bibfield  {title} {\bibinfo {title} {Nanoscale magnetometry with nv centers in diamond},\ }\href@noop {} {\bibfield  {journal} {\bibinfo  {journal} {MRS bulletin}\ }\textbf {\bibinfo {volume} {38}},\ \bibinfo {pages} {155} (\bibinfo {year} {2013})}\BibitemShut {NoStop}%
\bibitem [{\citenamefont {Zhang}\ \emph {et~al.}(2017)\citenamefont {Zhang}, \citenamefont {Pagano}, \citenamefont {Hess}, \citenamefont {Kyprianidis}, \citenamefont {Becker}, \citenamefont {Kaplan}, \citenamefont {Gorshkov}, \citenamefont {Gong},\ and\ \citenamefont {Monroe}}]{zhang2017observation}%
  \BibitemOpen
  \bibfield  {author} {\bibinfo {author} {\bibfnamefont {J.}~\bibnamefont {Zhang}}, \bibinfo {author} {\bibfnamefont {G.}~\bibnamefont {Pagano}}, \bibinfo {author} {\bibfnamefont {P.~W.}\ \bibnamefont {Hess}}, \bibinfo {author} {\bibfnamefont {A.}~\bibnamefont {Kyprianidis}}, \bibinfo {author} {\bibfnamefont {P.}~\bibnamefont {Becker}}, \bibinfo {author} {\bibfnamefont {H.}~\bibnamefont {Kaplan}}, \bibinfo {author} {\bibfnamefont {A.~V.}\ \bibnamefont {Gorshkov}}, \bibinfo {author} {\bibfnamefont {Z.-X.}\ \bibnamefont {Gong}},\ and\ \bibinfo {author} {\bibfnamefont {C.}~\bibnamefont {Monroe}},\ }\bibfield  {title} {\bibinfo {title} {Observation of a many-body dynamical phase transition with a 53-qubit quantum simulator},\ }\href@noop {} {\bibfield  {journal} {\bibinfo  {journal} {Nature}\ }\textbf {\bibinfo {volume} {551}},\ \bibinfo {pages} {601} (\bibinfo {year} {2017})}\BibitemShut {NoStop}%
\bibitem [{\citenamefont {Sch{\"a}fer}\ \emph {et~al.}(2020)\citenamefont {Sch{\"a}fer}, \citenamefont {Fukuhara}, \citenamefont {Sugawa}, \citenamefont {Takasu},\ and\ \citenamefont {Takahashi}}]{schafer2020tools}%
  \BibitemOpen
  \bibfield  {author} {\bibinfo {author} {\bibfnamefont {F.}~\bibnamefont {Sch{\"a}fer}}, \bibinfo {author} {\bibfnamefont {T.}~\bibnamefont {Fukuhara}}, \bibinfo {author} {\bibfnamefont {S.}~\bibnamefont {Sugawa}}, \bibinfo {author} {\bibfnamefont {Y.}~\bibnamefont {Takasu}},\ and\ \bibinfo {author} {\bibfnamefont {Y.}~\bibnamefont {Takahashi}},\ }\bibfield  {title} {\bibinfo {title} {Tools for quantum simulation with ultracold atoms in optical lattices},\ }\href@noop {} {\bibfield  {journal} {\bibinfo  {journal} {Nature Reviews Physics}\ }\textbf {\bibinfo {volume} {2}},\ \bibinfo {pages} {411} (\bibinfo {year} {2020})}\BibitemShut {NoStop}%
\bibitem [{\citenamefont {Essen}\ and\ \citenamefont {Parry}(1955)}]{essen1955atomic}%
  \BibitemOpen
  \bibfield  {author} {\bibinfo {author} {\bibfnamefont {L.}~\bibnamefont {Essen}}\ and\ \bibinfo {author} {\bibfnamefont {J.~V.}\ \bibnamefont {Parry}},\ }\bibfield  {title} {\bibinfo {title} {An atomic standard of frequency and time interval: a caesium resonator},\ }\href@noop {} {\bibfield  {journal} {\bibinfo  {journal} {Nature}\ }\textbf {\bibinfo {volume} {176}},\ \bibinfo {pages} {280} (\bibinfo {year} {1955})}\BibitemShut {NoStop}%
\bibitem [{\citenamefont {Aeppli}\ \emph {et~al.}(2024)\citenamefont {Aeppli}, \citenamefont {Kim}, \citenamefont {Warfield}, \citenamefont {Safronova},\ and\ \citenamefont {Ye}}]{aeppli2024clock}%
  \BibitemOpen
  \bibfield  {author} {\bibinfo {author} {\bibfnamefont {A.}~\bibnamefont {Aeppli}}, \bibinfo {author} {\bibfnamefont {K.}~\bibnamefont {Kim}}, \bibinfo {author} {\bibfnamefont {W.}~\bibnamefont {Warfield}}, \bibinfo {author} {\bibfnamefont {M.~S.}\ \bibnamefont {Safronova}},\ and\ \bibinfo {author} {\bibfnamefont {J.}~\bibnamefont {Ye}},\ }\bibfield  {title} {\bibinfo {title} {Clock with 8$\times$ 10-19 systematic uncertainty},\ }\href@noop {} {\bibfield  {journal} {\bibinfo  {journal} {Physical Review Letters}\ }\textbf {\bibinfo {volume} {133}},\ \bibinfo {pages} {023401} (\bibinfo {year} {2024})}\BibitemShut {NoStop}%
\bibitem [{\citenamefont {Jarmola}\ \emph {et~al.}(2021)\citenamefont {Jarmola}, \citenamefont {Lourette}, \citenamefont {Acosta}, \citenamefont {Birdwell}, \citenamefont {Bl{\"u}mler}, \citenamefont {Budker}, \citenamefont {Ivanov},\ and\ \citenamefont {Malinovsky}}]{jarmola2021demonstration}%
  \BibitemOpen
  \bibfield  {author} {\bibinfo {author} {\bibfnamefont {A.}~\bibnamefont {Jarmola}}, \bibinfo {author} {\bibfnamefont {S.}~\bibnamefont {Lourette}}, \bibinfo {author} {\bibfnamefont {V.~M.}\ \bibnamefont {Acosta}}, \bibinfo {author} {\bibfnamefont {A.~G.}\ \bibnamefont {Birdwell}}, \bibinfo {author} {\bibfnamefont {P.}~\bibnamefont {Bl{\"u}mler}}, \bibinfo {author} {\bibfnamefont {D.}~\bibnamefont {Budker}}, \bibinfo {author} {\bibfnamefont {T.}~\bibnamefont {Ivanov}},\ and\ \bibinfo {author} {\bibfnamefont {V.~S.}\ \bibnamefont {Malinovsky}},\ }\bibfield  {title} {\bibinfo {title} {Demonstration of diamond nuclear spin gyroscope},\ }\href@noop {} {\bibfield  {journal} {\bibinfo  {journal} {Science advances}\ }\textbf {\bibinfo {volume} {7}},\ \bibinfo {pages} {eabl3840} (\bibinfo {year} {2021})}\BibitemShut {NoStop}%
\bibitem [{\citenamefont {Soshenko}\ \emph {et~al.}(2021)\citenamefont {Soshenko}, \citenamefont {Bolshedvorskii}, \citenamefont {Rubinas}, \citenamefont {Sorokin}, \citenamefont {Smolyaninov}, \citenamefont {Vorobyov},\ and\ \citenamefont {Akimov}}]{soshenko2021nuclear}%
  \BibitemOpen
  \bibfield  {author} {\bibinfo {author} {\bibfnamefont {V.~V.}\ \bibnamefont {Soshenko}}, \bibinfo {author} {\bibfnamefont {S.~V.}\ \bibnamefont {Bolshedvorskii}}, \bibinfo {author} {\bibfnamefont {O.}~\bibnamefont {Rubinas}}, \bibinfo {author} {\bibfnamefont {V.~N.}\ \bibnamefont {Sorokin}}, \bibinfo {author} {\bibfnamefont {A.~N.}\ \bibnamefont {Smolyaninov}}, \bibinfo {author} {\bibfnamefont {V.~V.}\ \bibnamefont {Vorobyov}},\ and\ \bibinfo {author} {\bibfnamefont {A.~V.}\ \bibnamefont {Akimov}},\ }\bibfield  {title} {\bibinfo {title} {Nuclear spin gyroscope based on the nitrogen vacancy center in diamond},\ }\href@noop {} {\bibfield  {journal} {\bibinfo  {journal} {Physical Review Letters}\ }\textbf {\bibinfo {volume} {126}},\ \bibinfo {pages} {197702} (\bibinfo {year} {2021})}\BibitemShut {NoStop}%
\bibitem [{\citenamefont {Dicke}(1954)}]{dicke1954coherence}%
  \BibitemOpen
  \bibfield  {author} {\bibinfo {author} {\bibfnamefont {R.~H.}\ \bibnamefont {Dicke}},\ }\bibfield  {title} {\bibinfo {title} {Coherence in spontaneous radiation processes},\ }\href@noop {} {\bibfield  {journal} {\bibinfo  {journal} {Physical review}\ }\textbf {\bibinfo {volume} {93}},\ \bibinfo {pages} {99} (\bibinfo {year} {1954})}\BibitemShut {NoStop}%
\bibitem [{\citenamefont {Rose}\ \emph {et~al.}(2017)\citenamefont {Rose}, \citenamefont {Tyryshkin}, \citenamefont {Riemann}, \citenamefont {Abrosimov}, \citenamefont {Becker}, \citenamefont {Pohl}, \citenamefont {Thewalt}, \citenamefont {Itoh},\ and\ \citenamefont {Lyon}}]{rose2017coherent}%
  \BibitemOpen
  \bibfield  {author} {\bibinfo {author} {\bibfnamefont {B.}~\bibnamefont {Rose}}, \bibinfo {author} {\bibfnamefont {A.}~\bibnamefont {Tyryshkin}}, \bibinfo {author} {\bibfnamefont {H.}~\bibnamefont {Riemann}}, \bibinfo {author} {\bibfnamefont {N.}~\bibnamefont {Abrosimov}}, \bibinfo {author} {\bibfnamefont {P.}~\bibnamefont {Becker}}, \bibinfo {author} {\bibfnamefont {H.-J.}\ \bibnamefont {Pohl}}, \bibinfo {author} {\bibfnamefont {M.}~\bibnamefont {Thewalt}}, \bibinfo {author} {\bibfnamefont {K.~M.}\ \bibnamefont {Itoh}},\ and\ \bibinfo {author} {\bibfnamefont {S.~A.}\ \bibnamefont {Lyon}},\ }\bibfield  {title} {\bibinfo {title} {Coherent rabi dynamics of a superradiant spin ensemble in a microwave cavity},\ }\href@noop {} {\bibfield  {journal} {\bibinfo  {journal} {Physical Review X}\ }\textbf {\bibinfo {volume} {7}},\ \bibinfo {pages} {031002} (\bibinfo {year} {2017})}\BibitemShut {NoStop}%
\bibitem [{\citenamefont {Angerer}\ \emph {et~al.}(2018)\citenamefont {Angerer}, \citenamefont {Streltsov}, \citenamefont {Astner}, \citenamefont {Putz}, \citenamefont {Sumiya}, \citenamefont {Onoda}, \citenamefont {Isoya}, \citenamefont {Munro}, \citenamefont {Nemoto}, \citenamefont {Schmiedmayer} \emph {et~al.}}]{angerer2018superradiant}%
  \BibitemOpen
  \bibfield  {author} {\bibinfo {author} {\bibfnamefont {A.}~\bibnamefont {Angerer}}, \bibinfo {author} {\bibfnamefont {K.}~\bibnamefont {Streltsov}}, \bibinfo {author} {\bibfnamefont {T.}~\bibnamefont {Astner}}, \bibinfo {author} {\bibfnamefont {S.}~\bibnamefont {Putz}}, \bibinfo {author} {\bibfnamefont {H.}~\bibnamefont {Sumiya}}, \bibinfo {author} {\bibfnamefont {S.}~\bibnamefont {Onoda}}, \bibinfo {author} {\bibfnamefont {J.}~\bibnamefont {Isoya}}, \bibinfo {author} {\bibfnamefont {W.~J.}\ \bibnamefont {Munro}}, \bibinfo {author} {\bibfnamefont {K.}~\bibnamefont {Nemoto}}, \bibinfo {author} {\bibfnamefont {J.}~\bibnamefont {Schmiedmayer}}, \emph {et~al.},\ }\bibfield  {title} {\bibinfo {title} {Superradiant emission from colour centres in diamond},\ }\href@noop {} {\bibfield  {journal} {\bibinfo  {journal} {Nature Physics}\ }\textbf {\bibinfo {volume} {14}},\ \bibinfo {pages} {1168} (\bibinfo {year} {2018})}\BibitemShut {NoStop}%
\bibitem [{\citenamefont {Guerin}\ \emph {et~al.}(2016)\citenamefont {Guerin}, \citenamefont {Ara{\'u}jo},\ and\ \citenamefont {Kaiser}}]{guerin2016subradiance}%
  \BibitemOpen
  \bibfield  {author} {\bibinfo {author} {\bibfnamefont {W.}~\bibnamefont {Guerin}}, \bibinfo {author} {\bibfnamefont {M.~O.}\ \bibnamefont {Ara{\'u}jo}},\ and\ \bibinfo {author} {\bibfnamefont {R.}~\bibnamefont {Kaiser}},\ }\bibfield  {title} {\bibinfo {title} {Subradiance in a large cloud of cold atoms},\ }\href@noop {} {\bibfield  {journal} {\bibinfo  {journal} {Physical review letters}\ }\textbf {\bibinfo {volume} {116}},\ \bibinfo {pages} {083601} (\bibinfo {year} {2016})}\BibitemShut {NoStop}%
\bibitem [{\citenamefont {Weichselbaumer}\ \emph {et~al.}(2020)\citenamefont {Weichselbaumer}, \citenamefont {Zens}, \citenamefont {Zollitsch}, \citenamefont {Brandt}, \citenamefont {Rotter}, \citenamefont {Gross},\ and\ \citenamefont {Huebl}}]{weichselbaumer2020echo}%
  \BibitemOpen
  \bibfield  {author} {\bibinfo {author} {\bibfnamefont {S.}~\bibnamefont {Weichselbaumer}}, \bibinfo {author} {\bibfnamefont {M.}~\bibnamefont {Zens}}, \bibinfo {author} {\bibfnamefont {C.~W.}\ \bibnamefont {Zollitsch}}, \bibinfo {author} {\bibfnamefont {M.~S.}\ \bibnamefont {Brandt}}, \bibinfo {author} {\bibfnamefont {S.}~\bibnamefont {Rotter}}, \bibinfo {author} {\bibfnamefont {R.}~\bibnamefont {Gross}},\ and\ \bibinfo {author} {\bibfnamefont {H.}~\bibnamefont {Huebl}},\ }\bibfield  {title} {\bibinfo {title} {Echo trains in pulsed electron spin resonance of a strongly coupled spin ensemble},\ }\href@noop {} {\bibfield  {journal} {\bibinfo  {journal} {Physical Review Letters}\ }\textbf {\bibinfo {volume} {125}},\ \bibinfo {pages} {137701} (\bibinfo {year} {2020})}\BibitemShut {NoStop}%
\bibitem [{\citenamefont {Debnath}\ \emph {et~al.}(2020)\citenamefont {Debnath}, \citenamefont {Dold}, \citenamefont {Morton},\ and\ \citenamefont {M{\o}lmer}}]{debnath2020self}%
  \BibitemOpen
  \bibfield  {author} {\bibinfo {author} {\bibfnamefont {K.}~\bibnamefont {Debnath}}, \bibinfo {author} {\bibfnamefont {G.}~\bibnamefont {Dold}}, \bibinfo {author} {\bibfnamefont {J.~J.}\ \bibnamefont {Morton}},\ and\ \bibinfo {author} {\bibfnamefont {K.}~\bibnamefont {M{\o}lmer}},\ }\bibfield  {title} {\bibinfo {title} {Self-stimulated pulse echo trains from inhomogeneously broadened spin ensembles},\ }\href@noop {} {\bibfield  {journal} {\bibinfo  {journal} {Physical Review Letters}\ }\textbf {\bibinfo {volume} {125}},\ \bibinfo {pages} {137702} (\bibinfo {year} {2020})}\BibitemShut {NoStop}%
\bibitem [{\citenamefont {Bar-Gill}\ \emph {et~al.}(2013)\citenamefont {Bar-Gill}, \citenamefont {Pham}, \citenamefont {Jarmola}, \citenamefont {Budker},\ and\ \citenamefont {Walsworth}}]{bar2013solid}%
  \BibitemOpen
  \bibfield  {author} {\bibinfo {author} {\bibfnamefont {N.}~\bibnamefont {Bar-Gill}}, \bibinfo {author} {\bibfnamefont {L.~M.}\ \bibnamefont {Pham}}, \bibinfo {author} {\bibfnamefont {A.}~\bibnamefont {Jarmola}}, \bibinfo {author} {\bibfnamefont {D.}~\bibnamefont {Budker}},\ and\ \bibinfo {author} {\bibfnamefont {R.~L.}\ \bibnamefont {Walsworth}},\ }\bibfield  {title} {\bibinfo {title} {Solid-state electronic spin coherence time approaching one second},\ }\href@noop {} {\bibfield  {journal} {\bibinfo  {journal} {Nature communications}\ }\textbf {\bibinfo {volume} {4}},\ \bibinfo {pages} {1743} (\bibinfo {year} {2013})}\BibitemShut {NoStop}%
\bibitem [{\citenamefont {Wolfowicz}\ \emph {et~al.}(2013)\citenamefont {Wolfowicz}, \citenamefont {Tyryshkin}, \citenamefont {George}, \citenamefont {Riemann}, \citenamefont {Abrosimov}, \citenamefont {Becker}, \citenamefont {Pohl}, \citenamefont {Thewalt}, \citenamefont {Lyon},\ and\ \citenamefont {Morton}}]{wolfowicz2013atomic}%
  \BibitemOpen
  \bibfield  {author} {\bibinfo {author} {\bibfnamefont {G.}~\bibnamefont {Wolfowicz}}, \bibinfo {author} {\bibfnamefont {A.~M.}\ \bibnamefont {Tyryshkin}}, \bibinfo {author} {\bibfnamefont {R.~E.}\ \bibnamefont {George}}, \bibinfo {author} {\bibfnamefont {H.}~\bibnamefont {Riemann}}, \bibinfo {author} {\bibfnamefont {N.~V.}\ \bibnamefont {Abrosimov}}, \bibinfo {author} {\bibfnamefont {P.}~\bibnamefont {Becker}}, \bibinfo {author} {\bibfnamefont {H.-J.}\ \bibnamefont {Pohl}}, \bibinfo {author} {\bibfnamefont {M.~L.}\ \bibnamefont {Thewalt}}, \bibinfo {author} {\bibfnamefont {S.~A.}\ \bibnamefont {Lyon}},\ and\ \bibinfo {author} {\bibfnamefont {J.~J.}\ \bibnamefont {Morton}},\ }\bibfield  {title} {\bibinfo {title} {Atomic clock transitions in silicon-based spin qubits},\ }\href@noop {} {\bibfield  {journal} {\bibinfo  {journal} {Nature nanotechnology}\ }\textbf {\bibinfo {volume} {8}},\ \bibinfo {pages} {561} (\bibinfo {year} {2013})}\BibitemShut {NoStop}%
\bibitem [{\citenamefont {Tyryshkin}\ \emph {et~al.}(2012)\citenamefont {Tyryshkin}, \citenamefont {Tojo}, \citenamefont {Morton}, \citenamefont {Riemann}, \citenamefont {Abrosimov}, \citenamefont {Becker}, \citenamefont {Pohl}, \citenamefont {Schenkel}, \citenamefont {Thewalt}, \citenamefont {Itoh} \emph {et~al.}}]{tyryshkin2012electron}%
  \BibitemOpen
  \bibfield  {author} {\bibinfo {author} {\bibfnamefont {A.~M.}\ \bibnamefont {Tyryshkin}}, \bibinfo {author} {\bibfnamefont {S.}~\bibnamefont {Tojo}}, \bibinfo {author} {\bibfnamefont {J.~J.}\ \bibnamefont {Morton}}, \bibinfo {author} {\bibfnamefont {H.}~\bibnamefont {Riemann}}, \bibinfo {author} {\bibfnamefont {N.~V.}\ \bibnamefont {Abrosimov}}, \bibinfo {author} {\bibfnamefont {P.}~\bibnamefont {Becker}}, \bibinfo {author} {\bibfnamefont {H.-J.}\ \bibnamefont {Pohl}}, \bibinfo {author} {\bibfnamefont {T.}~\bibnamefont {Schenkel}}, \bibinfo {author} {\bibfnamefont {M.~L.}\ \bibnamefont {Thewalt}}, \bibinfo {author} {\bibfnamefont {K.~M.}\ \bibnamefont {Itoh}}, \emph {et~al.},\ }\bibfield  {title} {\bibinfo {title} {Electron spin coherence exceeding seconds in high-purity silicon},\ }\href@noop {} {\bibfield  {journal} {\bibinfo  {journal} {Nature materials}\ }\textbf {\bibinfo {volume} {11}},\ \bibinfo {pages} {143} (\bibinfo {year} {2012})}\BibitemShut {NoStop}%
\bibitem [{\citenamefont {Le~Dantec}\ \emph {et~al.}(2021)\citenamefont {Le~Dantec}, \citenamefont {Ran{\v{c}}i{\'c}}, \citenamefont {Lin}, \citenamefont {Billaud}, \citenamefont {Ranjan}, \citenamefont {Flanigan}, \citenamefont {Bertaina}, \citenamefont {Chaneli{\`e}re}, \citenamefont {Goldner}, \citenamefont {Erb} \emph {et~al.}}]{le2021twenty}%
  \BibitemOpen
  \bibfield  {author} {\bibinfo {author} {\bibfnamefont {M.}~\bibnamefont {Le~Dantec}}, \bibinfo {author} {\bibfnamefont {M.}~\bibnamefont {Ran{\v{c}}i{\'c}}}, \bibinfo {author} {\bibfnamefont {S.}~\bibnamefont {Lin}}, \bibinfo {author} {\bibfnamefont {E.}~\bibnamefont {Billaud}}, \bibinfo {author} {\bibfnamefont {V.}~\bibnamefont {Ranjan}}, \bibinfo {author} {\bibfnamefont {D.}~\bibnamefont {Flanigan}}, \bibinfo {author} {\bibfnamefont {S.}~\bibnamefont {Bertaina}}, \bibinfo {author} {\bibfnamefont {T.}~\bibnamefont {Chaneli{\`e}re}}, \bibinfo {author} {\bibfnamefont {P.}~\bibnamefont {Goldner}}, \bibinfo {author} {\bibfnamefont {A.}~\bibnamefont {Erb}}, \emph {et~al.},\ }\bibfield  {title} {\bibinfo {title} {Twenty-three--millisecond electron spin coherence of erbium ions in a natural-abundance crystal},\ }\href@noop {} {\bibfield  {journal} {\bibinfo  {journal} {Science advances}\ }\textbf {\bibinfo {volume} {7}},\ \bibinfo {pages} {eabj9786} (\bibinfo {year} {2021})}\BibitemShut {NoStop}%
\bibitem [{\citenamefont {Kubo}\ \emph {et~al.}(2010)\citenamefont {Kubo}, \citenamefont {Ong}, \citenamefont {Bertet}, \citenamefont {Vion}, \citenamefont {Jacques}, \citenamefont {Zheng}, \citenamefont {Dr{\'e}au}, \citenamefont {Roch}, \citenamefont {Auff{\`e}ves}, \citenamefont {Jelezko} \emph {et~al.}}]{kubo2010strong}%
  \BibitemOpen
  \bibfield  {author} {\bibinfo {author} {\bibfnamefont {Y.}~\bibnamefont {Kubo}}, \bibinfo {author} {\bibfnamefont {F.}~\bibnamefont {Ong}}, \bibinfo {author} {\bibfnamefont {P.}~\bibnamefont {Bertet}}, \bibinfo {author} {\bibfnamefont {D.}~\bibnamefont {Vion}}, \bibinfo {author} {\bibfnamefont {V.}~\bibnamefont {Jacques}}, \bibinfo {author} {\bibfnamefont {D.}~\bibnamefont {Zheng}}, \bibinfo {author} {\bibfnamefont {A.}~\bibnamefont {Dr{\'e}au}}, \bibinfo {author} {\bibfnamefont {J.-F.}\ \bibnamefont {Roch}}, \bibinfo {author} {\bibfnamefont {A.}~\bibnamefont {Auff{\`e}ves}}, \bibinfo {author} {\bibfnamefont {F.}~\bibnamefont {Jelezko}}, \emph {et~al.},\ }\bibfield  {title} {\bibinfo {title} {Strong coupling of a spin ensemble to a superconducting resonator},\ }\href@noop {} {\bibfield  {journal} {\bibinfo  {journal} {Physical review letters}\ }\textbf {\bibinfo {volume} {105}},\ \bibinfo {pages} {140502} (\bibinfo {year} {2010})}\BibitemShut {NoStop}%
\bibitem [{\citenamefont {Schuster}\ \emph {et~al.}(2010)\citenamefont {Schuster}, \citenamefont {Sears}, \citenamefont {Ginossar}, \citenamefont {DiCarlo}, \citenamefont {Frunzio}, \citenamefont {Morton}, \citenamefont {Wu}, \citenamefont {Briggs}, \citenamefont {Buckley}, \citenamefont {Awschalom} \emph {et~al.}}]{schuster2010high}%
  \BibitemOpen
  \bibfield  {author} {\bibinfo {author} {\bibfnamefont {D.}~\bibnamefont {Schuster}}, \bibinfo {author} {\bibfnamefont {A.}~\bibnamefont {Sears}}, \bibinfo {author} {\bibfnamefont {E.}~\bibnamefont {Ginossar}}, \bibinfo {author} {\bibfnamefont {L.}~\bibnamefont {DiCarlo}}, \bibinfo {author} {\bibfnamefont {L.}~\bibnamefont {Frunzio}}, \bibinfo {author} {\bibfnamefont {J.}~\bibnamefont {Morton}}, \bibinfo {author} {\bibfnamefont {H.}~\bibnamefont {Wu}}, \bibinfo {author} {\bibfnamefont {G.}~\bibnamefont {Briggs}}, \bibinfo {author} {\bibfnamefont {B.}~\bibnamefont {Buckley}}, \bibinfo {author} {\bibfnamefont {D.}~\bibnamefont {Awschalom}}, \emph {et~al.},\ }\bibfield  {title} {\bibinfo {title} {High-cooperativity coupling of electron-spin ensembles to superconducting cavities},\ }\href@noop {} {\bibfield  {journal} {\bibinfo  {journal} {Physical review letters}\ }\textbf {\bibinfo {volume} {105}},\ \bibinfo {pages} {140501} (\bibinfo {year} {2010})}\BibitemShut {NoStop}%
\bibitem [{\citenamefont {Ams{\"u}ss}\ \emph {et~al.}(2011)\citenamefont {Ams{\"u}ss}, \citenamefont {Koller}, \citenamefont {N{\"o}bauer}, \citenamefont {Putz}, \citenamefont {Rotter}, \citenamefont {Sandner}, \citenamefont {Schneider}, \citenamefont {Schramb{\"o}ck}, \citenamefont {Steinhauser}, \citenamefont {Ritsch} \emph {et~al.}}]{amsuss2011cavity}%
  \BibitemOpen
  \bibfield  {author} {\bibinfo {author} {\bibfnamefont {R.}~\bibnamefont {Ams{\"u}ss}}, \bibinfo {author} {\bibfnamefont {C.}~\bibnamefont {Koller}}, \bibinfo {author} {\bibfnamefont {T.}~\bibnamefont {N{\"o}bauer}}, \bibinfo {author} {\bibfnamefont {S.}~\bibnamefont {Putz}}, \bibinfo {author} {\bibfnamefont {S.}~\bibnamefont {Rotter}}, \bibinfo {author} {\bibfnamefont {K.}~\bibnamefont {Sandner}}, \bibinfo {author} {\bibfnamefont {S.}~\bibnamefont {Schneider}}, \bibinfo {author} {\bibfnamefont {M.}~\bibnamefont {Schramb{\"o}ck}}, \bibinfo {author} {\bibfnamefont {G.}~\bibnamefont {Steinhauser}}, \bibinfo {author} {\bibfnamefont {H.}~\bibnamefont {Ritsch}}, \emph {et~al.},\ }\bibfield  {title} {\bibinfo {title} {Cavity qed with magnetically coupled collective spin states},\ }\href@noop {} {\bibfield  {journal} {\bibinfo  {journal} {Physical review letters}\ }\textbf {\bibinfo {volume} {107}},\ \bibinfo {pages} {060502} (\bibinfo {year} {2011})}\BibitemShut {NoStop}%
\bibitem [{\citenamefont {Abe}\ \emph {et~al.}(2011)\citenamefont {Abe}, \citenamefont {Wu}, \citenamefont {Ardavan},\ and\ \citenamefont {Morton}}]{abe2011electron}%
  \BibitemOpen
  \bibfield  {author} {\bibinfo {author} {\bibfnamefont {E.}~\bibnamefont {Abe}}, \bibinfo {author} {\bibfnamefont {H.}~\bibnamefont {Wu}}, \bibinfo {author} {\bibfnamefont {A.}~\bibnamefont {Ardavan}},\ and\ \bibinfo {author} {\bibfnamefont {J.~J.}\ \bibnamefont {Morton}},\ }\bibfield  {title} {\bibinfo {title} {Electron spin ensemble strongly coupled to a three-dimensional microwave cavity},\ }\href@noop {} {\bibfield  {journal} {\bibinfo  {journal} {Applied Physics Letters}\ }\textbf {\bibinfo {volume} {98}} (\bibinfo {year} {2011})}\BibitemShut {NoStop}%
\bibitem [{\citenamefont {Sandner}\ \emph {et~al.}(2012)\citenamefont {Sandner}, \citenamefont {Ritsch}, \citenamefont {Ams{\"u}ss}, \citenamefont {Koller}, \citenamefont {N{\"o}bauer}, \citenamefont {Putz}, \citenamefont {Schmiedmayer},\ and\ \citenamefont {Majer}}]{sandner2012strong}%
  \BibitemOpen
  \bibfield  {author} {\bibinfo {author} {\bibfnamefont {K.}~\bibnamefont {Sandner}}, \bibinfo {author} {\bibfnamefont {H.}~\bibnamefont {Ritsch}}, \bibinfo {author} {\bibfnamefont {R.}~\bibnamefont {Ams{\"u}ss}}, \bibinfo {author} {\bibfnamefont {C.}~\bibnamefont {Koller}}, \bibinfo {author} {\bibfnamefont {T.}~\bibnamefont {N{\"o}bauer}}, \bibinfo {author} {\bibfnamefont {S.}~\bibnamefont {Putz}}, \bibinfo {author} {\bibfnamefont {J.}~\bibnamefont {Schmiedmayer}},\ and\ \bibinfo {author} {\bibfnamefont {J.}~\bibnamefont {Majer}},\ }\bibfield  {title} {\bibinfo {title} {Strong magnetic coupling of an inhomogeneous nitrogen-vacancy ensemble to a cavity},\ }\href@noop {} {\bibfield  {journal} {\bibinfo  {journal} {Physical Review A—Atomic, Molecular, and Optical Physics}\ }\textbf {\bibinfo {volume} {85}},\ \bibinfo {pages} {053806} (\bibinfo {year} {2012})}\BibitemShut {NoStop}%
\bibitem [{\citenamefont {Ranjan}\ \emph {et~al.}(2013)\citenamefont {Ranjan}, \citenamefont {De~Lange}, \citenamefont {Schutjens}, \citenamefont {Debelhoir}, \citenamefont {Groen}, \citenamefont {Szombati}, \citenamefont {Thoen}, \citenamefont {Klapwijk}, \citenamefont {Hanson},\ and\ \citenamefont {DiCarlo}}]{ranjan2013probing}%
  \BibitemOpen
  \bibfield  {author} {\bibinfo {author} {\bibfnamefont {V.}~\bibnamefont {Ranjan}}, \bibinfo {author} {\bibfnamefont {G.}~\bibnamefont {De~Lange}}, \bibinfo {author} {\bibfnamefont {R.}~\bibnamefont {Schutjens}}, \bibinfo {author} {\bibfnamefont {T.}~\bibnamefont {Debelhoir}}, \bibinfo {author} {\bibfnamefont {J.}~\bibnamefont {Groen}}, \bibinfo {author} {\bibfnamefont {D.}~\bibnamefont {Szombati}}, \bibinfo {author} {\bibfnamefont {D.}~\bibnamefont {Thoen}}, \bibinfo {author} {\bibfnamefont {T.}~\bibnamefont {Klapwijk}}, \bibinfo {author} {\bibfnamefont {R.}~\bibnamefont {Hanson}},\ and\ \bibinfo {author} {\bibfnamefont {L.}~\bibnamefont {DiCarlo}},\ }\bibfield  {title} {\bibinfo {title} {Probing dynamics of an electron-spin ensemble via a superconducting resonator},\ }\href@noop {} {\bibfield  {journal} {\bibinfo  {journal} {Physical Review Letters}\ }\textbf {\bibinfo {volume} {110}},\ \bibinfo {pages} {067004} (\bibinfo {year} {2013})}\BibitemShut {NoStop}%
\bibitem [{\citenamefont {Bushev}\ \emph {et~al.}(2011)\citenamefont {Bushev}, \citenamefont {Feofanov}, \citenamefont {Rotzinger}, \citenamefont {Protopopov}, \citenamefont {Cole}, \citenamefont {Wilson}, \citenamefont {Fischer}, \citenamefont {Lukashenko},\ and\ \citenamefont {Ustinov}}]{bushev2011ultralow}%
  \BibitemOpen
  \bibfield  {author} {\bibinfo {author} {\bibfnamefont {P.}~\bibnamefont {Bushev}}, \bibinfo {author} {\bibfnamefont {A.}~\bibnamefont {Feofanov}}, \bibinfo {author} {\bibfnamefont {H.}~\bibnamefont {Rotzinger}}, \bibinfo {author} {\bibfnamefont {I.}~\bibnamefont {Protopopov}}, \bibinfo {author} {\bibfnamefont {J.}~\bibnamefont {Cole}}, \bibinfo {author} {\bibfnamefont {C.}~\bibnamefont {Wilson}}, \bibinfo {author} {\bibfnamefont {G.}~\bibnamefont {Fischer}}, \bibinfo {author} {\bibfnamefont {A.}~\bibnamefont {Lukashenko}},\ and\ \bibinfo {author} {\bibfnamefont {A.}~\bibnamefont {Ustinov}},\ }\bibfield  {title} {\bibinfo {title} {Ultralow-power spectroscopy of a rare-earth spin ensemble using a superconducting resonator},\ }\href@noop {} {\bibfield  {journal} {\bibinfo  {journal} {Physical Review B—Condensed Matter and Materials Physics}\ }\textbf {\bibinfo {volume} {84}},\ \bibinfo {pages} {060501} (\bibinfo {year} {2011})}\BibitemShut {NoStop}%
\bibitem [{\citenamefont {O’Sullivan}\ \emph {et~al.}(2022)\citenamefont {O’Sullivan}, \citenamefont {Kennedy}, \citenamefont {Debnath}, \citenamefont {Alexander}, \citenamefont {Zollitsch}, \citenamefont {{\v{S}}im{\.e}nas}, \citenamefont {Hashim}, \citenamefont {Thomas}, \citenamefont {Withington}, \citenamefont {Siddiqi} \emph {et~al.}}]{o2022random}%
  \BibitemOpen
  \bibfield  {author} {\bibinfo {author} {\bibfnamefont {J.}~\bibnamefont {O’Sullivan}}, \bibinfo {author} {\bibfnamefont {O.~W.}\ \bibnamefont {Kennedy}}, \bibinfo {author} {\bibfnamefont {K.}~\bibnamefont {Debnath}}, \bibinfo {author} {\bibfnamefont {J.}~\bibnamefont {Alexander}}, \bibinfo {author} {\bibfnamefont {C.~W.}\ \bibnamefont {Zollitsch}}, \bibinfo {author} {\bibfnamefont {M.}~\bibnamefont {{\v{S}}im{\.e}nas}}, \bibinfo {author} {\bibfnamefont {A.}~\bibnamefont {Hashim}}, \bibinfo {author} {\bibfnamefont {C.~N.}\ \bibnamefont {Thomas}}, \bibinfo {author} {\bibfnamefont {S.}~\bibnamefont {Withington}}, \bibinfo {author} {\bibfnamefont {I.}~\bibnamefont {Siddiqi}}, \emph {et~al.},\ }\bibfield  {title} {\bibinfo {title} {Random-access quantum memory using chirped pulse phase encoding},\ }\href@noop {} {\bibfield  {journal} {\bibinfo  {journal} {Physical Review X}\ }\textbf {\bibinfo {volume} {12}},\ \bibinfo {pages} {041014} (\bibinfo {year} {2022})}\BibitemShut {NoStop}%
\bibitem [{\citenamefont {Ranjan}\ \emph {et~al.}(2022)\citenamefont {Ranjan}, \citenamefont {Wen}, \citenamefont {Keyser}, \citenamefont {Kubatkin}, \citenamefont {Danilov}, \citenamefont {Lindstr{\"o}m}, \citenamefont {Bertet},\ and\ \citenamefont {de~Graaf}}]{ranjan2022spin}%
  \BibitemOpen
  \bibfield  {author} {\bibinfo {author} {\bibfnamefont {V.}~\bibnamefont {Ranjan}}, \bibinfo {author} {\bibfnamefont {Y.}~\bibnamefont {Wen}}, \bibinfo {author} {\bibfnamefont {A.}~\bibnamefont {Keyser}}, \bibinfo {author} {\bibfnamefont {S.}~\bibnamefont {Kubatkin}}, \bibinfo {author} {\bibfnamefont {A.}~\bibnamefont {Danilov}}, \bibinfo {author} {\bibfnamefont {T.}~\bibnamefont {Lindstr{\"o}m}}, \bibinfo {author} {\bibfnamefont {P.}~\bibnamefont {Bertet}},\ and\ \bibinfo {author} {\bibfnamefont {S.}~\bibnamefont {de~Graaf}},\ }\bibfield  {title} {\bibinfo {title} {Spin-echo silencing using a current-biased frequency-tunable resonator},\ }\href@noop {} {\bibfield  {journal} {\bibinfo  {journal} {Physical Review Letters}\ }\textbf {\bibinfo {volume} {129}},\ \bibinfo {pages} {180504} (\bibinfo {year} {2022})}\BibitemShut {NoStop}%
\bibitem [{\citenamefont {Bauch}\ \emph {et~al.}(2018)\citenamefont {Bauch}, \citenamefont {Hart}, \citenamefont {Schloss}, \citenamefont {Turner}, \citenamefont {Barry}, \citenamefont {Kehayias}, \citenamefont {Singh},\ and\ \citenamefont {Walsworth}}]{bauch2018ultralong}%
  \BibitemOpen
  \bibfield  {author} {\bibinfo {author} {\bibfnamefont {E.}~\bibnamefont {Bauch}}, \bibinfo {author} {\bibfnamefont {C.~A.}\ \bibnamefont {Hart}}, \bibinfo {author} {\bibfnamefont {J.~M.}\ \bibnamefont {Schloss}}, \bibinfo {author} {\bibfnamefont {M.~J.}\ \bibnamefont {Turner}}, \bibinfo {author} {\bibfnamefont {J.~F.}\ \bibnamefont {Barry}}, \bibinfo {author} {\bibfnamefont {P.}~\bibnamefont {Kehayias}}, \bibinfo {author} {\bibfnamefont {S.}~\bibnamefont {Singh}},\ and\ \bibinfo {author} {\bibfnamefont {R.~L.}\ \bibnamefont {Walsworth}},\ }\bibfield  {title} {\bibinfo {title} {Ultralong dephasing times in solid-state spin ensembles via quantum control},\ }\href@noop {} {\bibfield  {journal} {\bibinfo  {journal} {Physical Review X}\ }\textbf {\bibinfo {volume} {8}},\ \bibinfo {pages} {031025} (\bibinfo {year} {2018})}\BibitemShut {NoStop}%
\bibitem [{\citenamefont {Bauch}\ \emph {et~al.}(2020)\citenamefont {Bauch}, \citenamefont {Singh}, \citenamefont {Lee}, \citenamefont {Hart}, \citenamefont {Schloss}, \citenamefont {Turner}, \citenamefont {Barry}, \citenamefont {Pham}, \citenamefont {Bar-Gill}, \citenamefont {Yelin} \emph {et~al.}}]{bauch2020decoherence}%
  \BibitemOpen
  \bibfield  {author} {\bibinfo {author} {\bibfnamefont {E.}~\bibnamefont {Bauch}}, \bibinfo {author} {\bibfnamefont {S.}~\bibnamefont {Singh}}, \bibinfo {author} {\bibfnamefont {J.}~\bibnamefont {Lee}}, \bibinfo {author} {\bibfnamefont {C.~A.}\ \bibnamefont {Hart}}, \bibinfo {author} {\bibfnamefont {J.~M.}\ \bibnamefont {Schloss}}, \bibinfo {author} {\bibfnamefont {M.~J.}\ \bibnamefont {Turner}}, \bibinfo {author} {\bibfnamefont {J.~F.}\ \bibnamefont {Barry}}, \bibinfo {author} {\bibfnamefont {L.~M.}\ \bibnamefont {Pham}}, \bibinfo {author} {\bibfnamefont {N.}~\bibnamefont {Bar-Gill}}, \bibinfo {author} {\bibfnamefont {S.~F.}\ \bibnamefont {Yelin}}, \emph {et~al.},\ }\bibfield  {title} {\bibinfo {title} {Decoherence of ensembles of nitrogen-vacancy centers in diamond},\ }\href@noop {} {\bibfield  {journal} {\bibinfo  {journal} {Physical Review B}\ }\textbf {\bibinfo {volume} {102}},\ \bibinfo {pages} {134210} (\bibinfo {year} {2020})}\BibitemShut {NoStop}%
\bibitem [{\citenamefont {Doherty}\ \emph {et~al.}(2011)\citenamefont {Doherty}, \citenamefont {Dolde}, \citenamefont {Fedder}, \citenamefont {Jelezko}, \citenamefont {Wrachtrup}, \citenamefont {Manson},\ and\ \citenamefont {Hollenberg}}]{doherty2011theory}%
  \BibitemOpen
  \bibfield  {author} {\bibinfo {author} {\bibfnamefont {M.}~\bibnamefont {Doherty}}, \bibinfo {author} {\bibfnamefont {F.}~\bibnamefont {Dolde}}, \bibinfo {author} {\bibfnamefont {H.}~\bibnamefont {Fedder}}, \bibinfo {author} {\bibfnamefont {F.}~\bibnamefont {Jelezko}}, \bibinfo {author} {\bibfnamefont {J.}~\bibnamefont {Wrachtrup}}, \bibinfo {author} {\bibfnamefont {N.}~\bibnamefont {Manson}},\ and\ \bibinfo {author} {\bibfnamefont {L.}~\bibnamefont {Hollenberg}},\ }\bibfield  {title} {\bibinfo {title} {Theory of the ground state spin of the nv-center in diamond: Ii. spin solutions, time-evolution, relaxation and inhomogeneous dephasing},\ }\href@noop {} {\bibfield  {journal} {\bibinfo  {journal} {arXiv preprint arXiv:1111.5882}\ } (\bibinfo {year} {2011})}\BibitemShut {NoStop}%
\bibitem [{\citenamefont {Jamonneau}\ \emph {et~al.}(2016)\citenamefont {Jamonneau}, \citenamefont {Lesik}, \citenamefont {Tetienne}, \citenamefont {Alvizu}, \citenamefont {Mayer}, \citenamefont {Dr{\'e}au}, \citenamefont {Kosen}, \citenamefont {Roch}, \citenamefont {Pezzagna}, \citenamefont {Meijer} \emph {et~al.}}]{jamonneau2016competition}%
  \BibitemOpen
  \bibfield  {author} {\bibinfo {author} {\bibfnamefont {P.}~\bibnamefont {Jamonneau}}, \bibinfo {author} {\bibfnamefont {M.}~\bibnamefont {Lesik}}, \bibinfo {author} {\bibfnamefont {J.}~\bibnamefont {Tetienne}}, \bibinfo {author} {\bibfnamefont {I.}~\bibnamefont {Alvizu}}, \bibinfo {author} {\bibfnamefont {L.}~\bibnamefont {Mayer}}, \bibinfo {author} {\bibfnamefont {A.}~\bibnamefont {Dr{\'e}au}}, \bibinfo {author} {\bibfnamefont {S.}~\bibnamefont {Kosen}}, \bibinfo {author} {\bibfnamefont {J.-F.}\ \bibnamefont {Roch}}, \bibinfo {author} {\bibfnamefont {S.}~\bibnamefont {Pezzagna}}, \bibinfo {author} {\bibfnamefont {J.}~\bibnamefont {Meijer}}, \emph {et~al.},\ }\bibfield  {title} {\bibinfo {title} {Competition between electric field and magnetic field noise in the decoherence of a single spin in diamond},\ }\href@noop {} {\bibfield  {journal} {\bibinfo  {journal} {Physical Review B}\ }\textbf {\bibinfo {volume} {93}},\ \bibinfo {pages} {024305} (\bibinfo {year} {2016})}\BibitemShut {NoStop}%
\bibitem [{\citenamefont {Ranjan}\ \emph {et~al.}(2020{\natexlab{b}})\citenamefont {Ranjan}, \citenamefont {Probst}, \citenamefont {Albanese}, \citenamefont {Schenkel}, \citenamefont {Vion}, \citenamefont {Esteve}, \citenamefont {Morton},\ and\ \citenamefont {Bertet}}]{ranjan2020electron}%
  \BibitemOpen
  \bibfield  {author} {\bibinfo {author} {\bibfnamefont {V.}~\bibnamefont {Ranjan}}, \bibinfo {author} {\bibfnamefont {S.}~\bibnamefont {Probst}}, \bibinfo {author} {\bibfnamefont {B.}~\bibnamefont {Albanese}}, \bibinfo {author} {\bibfnamefont {T.}~\bibnamefont {Schenkel}}, \bibinfo {author} {\bibfnamefont {D.}~\bibnamefont {Vion}}, \bibinfo {author} {\bibfnamefont {D.}~\bibnamefont {Esteve}}, \bibinfo {author} {\bibfnamefont {J.}~\bibnamefont {Morton}},\ and\ \bibinfo {author} {\bibfnamefont {P.}~\bibnamefont {Bertet}},\ }\bibfield  {title} {\bibinfo {title} {Electron spin resonance spectroscopy with femtoliter detection volume},\ }\href@noop {} {\bibfield  {journal} {\bibinfo  {journal} {Applied Physics Letters}\ }\textbf {\bibinfo {volume} {116}} (\bibinfo {year} {2020}{\natexlab{b}})}\BibitemShut {NoStop}%
\bibitem [{\citenamefont {Ranjan}\ \emph {et~al.}(2021)\citenamefont {Ranjan}, \citenamefont {Albanese}, \citenamefont {Albertinale}, \citenamefont {Billaud}, \citenamefont {Flanigan}, \citenamefont {Pla}, \citenamefont {Schenkel}, \citenamefont {Vion}, \citenamefont {Esteve}, \citenamefont {Flurin} \emph {et~al.}}]{ranjan2021spatially}%
  \BibitemOpen
  \bibfield  {author} {\bibinfo {author} {\bibfnamefont {V.}~\bibnamefont {Ranjan}}, \bibinfo {author} {\bibfnamefont {B.}~\bibnamefont {Albanese}}, \bibinfo {author} {\bibfnamefont {E.}~\bibnamefont {Albertinale}}, \bibinfo {author} {\bibfnamefont {E.}~\bibnamefont {Billaud}}, \bibinfo {author} {\bibfnamefont {D.}~\bibnamefont {Flanigan}}, \bibinfo {author} {\bibfnamefont {J.}~\bibnamefont {Pla}}, \bibinfo {author} {\bibfnamefont {T.}~\bibnamefont {Schenkel}}, \bibinfo {author} {\bibfnamefont {D.}~\bibnamefont {Vion}}, \bibinfo {author} {\bibfnamefont {D.}~\bibnamefont {Esteve}}, \bibinfo {author} {\bibfnamefont {E.}~\bibnamefont {Flurin}}, \emph {et~al.},\ }\bibfield  {title} {\bibinfo {title} {Spatially resolved decoherence of donor spins in silicon strained by a metallic electrode},\ }\href@noop {} {\bibfield  {journal} {\bibinfo  {journal} {Physical Review X}\ }\textbf {\bibinfo {volume} {11}},\ \bibinfo {pages} {031036} (\bibinfo {year} {2021})}\BibitemShut {NoStop}%
\bibitem [{\citenamefont {Pla}\ \emph {et~al.}(2018)\citenamefont {Pla}, \citenamefont {Bienfait}, \citenamefont {Pica}, \citenamefont {Mansir}, \citenamefont {Mohiyaddin}, \citenamefont {Zeng}, \citenamefont {Niquet}, \citenamefont {Morello}, \citenamefont {Schenkel}, \citenamefont {Morton} \emph {et~al.}}]{pla2018strain}%
  \BibitemOpen
  \bibfield  {author} {\bibinfo {author} {\bibfnamefont {J.}~\bibnamefont {Pla}}, \bibinfo {author} {\bibfnamefont {A.}~\bibnamefont {Bienfait}}, \bibinfo {author} {\bibfnamefont {G.}~\bibnamefont {Pica}}, \bibinfo {author} {\bibfnamefont {J.}~\bibnamefont {Mansir}}, \bibinfo {author} {\bibfnamefont {F.}~\bibnamefont {Mohiyaddin}}, \bibinfo {author} {\bibfnamefont {Z.}~\bibnamefont {Zeng}}, \bibinfo {author} {\bibfnamefont {Y.-M.}\ \bibnamefont {Niquet}}, \bibinfo {author} {\bibfnamefont {A.}~\bibnamefont {Morello}}, \bibinfo {author} {\bibfnamefont {T.}~\bibnamefont {Schenkel}}, \bibinfo {author} {\bibfnamefont {J.}~\bibnamefont {Morton}}, \emph {et~al.},\ }\bibfield  {title} {\bibinfo {title} {Strain-induced spin-resonance shifts in silicon devices},\ }\href@noop {} {\bibfield  {journal} {\bibinfo  {journal} {Physical Review Applied}\ }\textbf {\bibinfo {volume} {9}},\ \bibinfo {pages} {044014} (\bibinfo {year} {2018})}\BibitemShut {NoStop}%
\bibitem [{\citenamefont {Acosta}\ \emph {et~al.}(2010)\citenamefont {Acosta}, \citenamefont {Bauch}, \citenamefont {Ledbetter}, \citenamefont {Waxman}, \citenamefont {Bouchard},\ and\ \citenamefont {Budker}}]{acosta2010temperature}%
  \BibitemOpen
  \bibfield  {author} {\bibinfo {author} {\bibfnamefont {V.~M.}\ \bibnamefont {Acosta}}, \bibinfo {author} {\bibfnamefont {E.}~\bibnamefont {Bauch}}, \bibinfo {author} {\bibfnamefont {M.~P.}\ \bibnamefont {Ledbetter}}, \bibinfo {author} {\bibfnamefont {A.}~\bibnamefont {Waxman}}, \bibinfo {author} {\bibfnamefont {L.-S.}\ \bibnamefont {Bouchard}},\ and\ \bibinfo {author} {\bibfnamefont {D.}~\bibnamefont {Budker}},\ }\bibfield  {title} {\bibinfo {title} {Temperature dependence of the nitrogen-vacancy magnetic resonance in diamond},\ }\href@noop {} {\bibfield  {journal} {\bibinfo  {journal} {Physical review letters}\ }\textbf {\bibinfo {volume} {104}},\ \bibinfo {pages} {070801} (\bibinfo {year} {2010})}\BibitemShut {NoStop}%
\bibitem [{\citenamefont {Toyli}\ \emph {et~al.}(2013)\citenamefont {Toyli}, \citenamefont {de~Las~Casas}, \citenamefont {Christle}, \citenamefont {Dobrovitski},\ and\ \citenamefont {Awschalom}}]{toyli2013fluorescence}%
  \BibitemOpen
  \bibfield  {author} {\bibinfo {author} {\bibfnamefont {D.~M.}\ \bibnamefont {Toyli}}, \bibinfo {author} {\bibfnamefont {C.~F.}\ \bibnamefont {de~Las~Casas}}, \bibinfo {author} {\bibfnamefont {D.~J.}\ \bibnamefont {Christle}}, \bibinfo {author} {\bibfnamefont {V.~V.}\ \bibnamefont {Dobrovitski}},\ and\ \bibinfo {author} {\bibfnamefont {D.~D.}\ \bibnamefont {Awschalom}},\ }\bibfield  {title} {\bibinfo {title} {Fluorescence thermometry enhanced by the quantum coherence of single spins in diamond},\ }\href@noop {} {\bibfield  {journal} {\bibinfo  {journal} {Proceedings of the National Academy of Sciences}\ }\textbf {\bibinfo {volume} {110}},\ \bibinfo {pages} {8417} (\bibinfo {year} {2013})}\BibitemShut {NoStop}%
\bibitem [{\citenamefont {Gr{\`e}zes}(2015)}]{grezes2015thesis}%
  \BibitemOpen
  \bibfield  {author} {\bibinfo {author} {\bibfnamefont {C.}~\bibnamefont {Gr{\`e}zes}},\ }\emph {\bibinfo {title} {Towards a spin ensemble quantum memory for superconducting qubits}},\ \href@noop {} {Ph.D. thesis},\ \bibinfo  {school} {Universit{\'e} Pierre et Marie Curie (Paris VI)} (\bibinfo {year} {2015})\BibitemShut {NoStop}%
\bibitem [{\citenamefont {Putz}\ \emph {et~al.}(2014)\citenamefont {Putz}, \citenamefont {Krimer}, \citenamefont {Ams{\"u}ss}, \citenamefont {Valookaran}, \citenamefont {N{\"o}bauer}, \citenamefont {Schmiedmayer}, \citenamefont {Rotter},\ and\ \citenamefont {Majer}}]{putz2014protecting}%
  \BibitemOpen
  \bibfield  {author} {\bibinfo {author} {\bibfnamefont {S.}~\bibnamefont {Putz}}, \bibinfo {author} {\bibfnamefont {D.~O.}\ \bibnamefont {Krimer}}, \bibinfo {author} {\bibfnamefont {R.}~\bibnamefont {Ams{\"u}ss}}, \bibinfo {author} {\bibfnamefont {A.}~\bibnamefont {Valookaran}}, \bibinfo {author} {\bibfnamefont {T.}~\bibnamefont {N{\"o}bauer}}, \bibinfo {author} {\bibfnamefont {J.}~\bibnamefont {Schmiedmayer}}, \bibinfo {author} {\bibfnamefont {S.}~\bibnamefont {Rotter}},\ and\ \bibinfo {author} {\bibfnamefont {J.}~\bibnamefont {Majer}},\ }\bibfield  {title} {\bibinfo {title} {Protecting a spin ensemble against decoherence in the strong-coupling regime of cavity qed},\ }\href@noop {} {\bibfield  {journal} {\bibinfo  {journal} {Nature Physics}\ }\textbf {\bibinfo {volume} {10}},\ \bibinfo {pages} {720} (\bibinfo {year} {2014})}\BibitemShut {NoStop}%
\bibitem [{\citenamefont {Krimer}\ \emph {et~al.}(2015)\citenamefont {Krimer}, \citenamefont {Hartl},\ and\ \citenamefont {Rotter}}]{krimer2015hybrid}%
  \BibitemOpen
  \bibfield  {author} {\bibinfo {author} {\bibfnamefont {D.~O.}\ \bibnamefont {Krimer}}, \bibinfo {author} {\bibfnamefont {B.}~\bibnamefont {Hartl}},\ and\ \bibinfo {author} {\bibfnamefont {S.}~\bibnamefont {Rotter}},\ }\bibfield  {title} {\bibinfo {title} {Hybrid quantum systems with collectively coupled spin states: suppression of decoherence through spectral hole burning},\ }\href@noop {} {\bibfield  {journal} {\bibinfo  {journal} {Physical Review Letters}\ }\textbf {\bibinfo {volume} {115}},\ \bibinfo {pages} {033601} (\bibinfo {year} {2015})}\BibitemShut {NoStop}%
\bibitem [{\citenamefont {Putz}\ \emph {et~al.}(2017)\citenamefont {Putz}, \citenamefont {Angerer}, \citenamefont {Krimer}, \citenamefont {Glattauer}, \citenamefont {Munro}, \citenamefont {Rotter}, \citenamefont {Schmiedmayer},\ and\ \citenamefont {Majer}}]{putz2017spectral}%
  \BibitemOpen
  \bibfield  {author} {\bibinfo {author} {\bibfnamefont {S.}~\bibnamefont {Putz}}, \bibinfo {author} {\bibfnamefont {A.}~\bibnamefont {Angerer}}, \bibinfo {author} {\bibfnamefont {D.~O.}\ \bibnamefont {Krimer}}, \bibinfo {author} {\bibfnamefont {R.}~\bibnamefont {Glattauer}}, \bibinfo {author} {\bibfnamefont {W.~J.}\ \bibnamefont {Munro}}, \bibinfo {author} {\bibfnamefont {S.}~\bibnamefont {Rotter}}, \bibinfo {author} {\bibfnamefont {J.}~\bibnamefont {Schmiedmayer}},\ and\ \bibinfo {author} {\bibfnamefont {J.}~\bibnamefont {Majer}},\ }\bibfield  {title} {\bibinfo {title} {Spectral hole burning and its application in microwave photonics},\ }\href@noop {} {\bibfield  {journal} {\bibinfo  {journal} {Nature Photonics}\ }\textbf {\bibinfo {volume} {11}},\ \bibinfo {pages} {36} (\bibinfo {year} {2017})}\BibitemShut {NoStop}%
\bibitem [{\citenamefont {Tavis}\ and\ \citenamefont {Cummings}(1968)}]{tavis1968exact}%
  \BibitemOpen
  \bibfield  {author} {\bibinfo {author} {\bibfnamefont {M.}~\bibnamefont {Tavis}}\ and\ \bibinfo {author} {\bibfnamefont {F.~W.}\ \bibnamefont {Cummings}},\ }\bibfield  {title} {\bibinfo {title} {Exact solution for an n-molecule—radiation-field hamiltonian},\ }\href@noop {} {\bibfield  {journal} {\bibinfo  {journal} {Physical Review}\ }\textbf {\bibinfo {volume} {170}},\ \bibinfo {pages} {379} (\bibinfo {year} {1968})}\BibitemShut {NoStop}%
\bibitem [{\citenamefont {Gardiner}\ and\ \citenamefont {Collett}(1985)}]{gardiner1985input}%
  \BibitemOpen
  \bibfield  {author} {\bibinfo {author} {\bibfnamefont {C.~W.}\ \bibnamefont {Gardiner}}\ and\ \bibinfo {author} {\bibfnamefont {M.~J.}\ \bibnamefont {Collett}},\ }\bibfield  {title} {\bibinfo {title} {Input and output in damped quantum systems: Quantum stochastic differential equations and the master equation},\ }\href@noop {} {\bibfield  {journal} {\bibinfo  {journal} {Physical Review A}\ }\textbf {\bibinfo {volume} {31}},\ \bibinfo {pages} {3761} (\bibinfo {year} {1985})}\BibitemShut {NoStop}%
\bibitem [{\citenamefont {Probst}\ \emph {et~al.}(2014)\citenamefont {Probst}, \citenamefont {Tkal{\v{c}}ec}, \citenamefont {Rotzinger}, \citenamefont {Rieger}, \citenamefont {Le~Floch}, \citenamefont {Goryachev}, \citenamefont {Tobar}, \citenamefont {Ustinov},\ and\ \citenamefont {Bushev}}]{probst2014three}%
  \BibitemOpen
  \bibfield  {author} {\bibinfo {author} {\bibfnamefont {S.}~\bibnamefont {Probst}}, \bibinfo {author} {\bibfnamefont {A.}~\bibnamefont {Tkal{\v{c}}ec}}, \bibinfo {author} {\bibfnamefont {H.}~\bibnamefont {Rotzinger}}, \bibinfo {author} {\bibfnamefont {D.}~\bibnamefont {Rieger}}, \bibinfo {author} {\bibfnamefont {J.-M.}\ \bibnamefont {Le~Floch}}, \bibinfo {author} {\bibfnamefont {M.}~\bibnamefont {Goryachev}}, \bibinfo {author} {\bibfnamefont {M.~E.}\ \bibnamefont {Tobar}}, \bibinfo {author} {\bibfnamefont {A.~V.}\ \bibnamefont {Ustinov}},\ and\ \bibinfo {author} {\bibfnamefont {P.~A.}\ \bibnamefont {Bushev}},\ }\bibfield  {title} {\bibinfo {title} {Three-dimensional cavity quantum electrodynamics with a rare-earth spin ensemble},\ }\href@noop {} {\bibfield  {journal} {\bibinfo  {journal} {Physical Review B}\ }\textbf {\bibinfo {volume} {90}},\ \bibinfo {pages} {100404} (\bibinfo {year} {2014})}\BibitemShut {NoStop}%
\bibitem [{\citenamefont {Tobar}\ \emph {et~al.}(1998)\citenamefont {Tobar}, \citenamefont {Krupka}, \citenamefont {Ivanov},\ and\ \citenamefont {Woode}}]{tobar1998anisotropic}%
  \BibitemOpen
  \bibfield  {author} {\bibinfo {author} {\bibfnamefont {M.~E.}\ \bibnamefont {Tobar}}, \bibinfo {author} {\bibfnamefont {J.}~\bibnamefont {Krupka}}, \bibinfo {author} {\bibfnamefont {E.~N.}\ \bibnamefont {Ivanov}},\ and\ \bibinfo {author} {\bibfnamefont {R.~A.}\ \bibnamefont {Woode}},\ }\bibfield  {title} {\bibinfo {title} {Anisotropic complex permittivity measurements of mono-crystalline rutile between 10 and 300 k},\ }\href@noop {} {\bibfield  {journal} {\bibinfo  {journal} {Journal of Applied Physics}\ }\textbf {\bibinfo {volume} {83}},\ \bibinfo {pages} {1604} (\bibinfo {year} {1998})}\BibitemShut {NoStop}%
\bibitem [{\citenamefont {Kato}\ \emph {et~al.}(2023)\citenamefont {Kato}, \citenamefont {Sasaki}, \citenamefont {Matsuura}, \citenamefont {Usami},\ and\ \citenamefont {Nakamura}}]{kato2023high}%
  \BibitemOpen
  \bibfield  {author} {\bibinfo {author} {\bibfnamefont {K.}~\bibnamefont {Kato}}, \bibinfo {author} {\bibfnamefont {R.}~\bibnamefont {Sasaki}}, \bibinfo {author} {\bibfnamefont {K.}~\bibnamefont {Matsuura}}, \bibinfo {author} {\bibfnamefont {K.}~\bibnamefont {Usami}},\ and\ \bibinfo {author} {\bibfnamefont {Y.}~\bibnamefont {Nakamura}},\ }\bibfield  {title} {\bibinfo {title} {High-cooperativity cavity magnon-polariton using a high-q dielectric resonator},\ }\href@noop {} {\bibfield  {journal} {\bibinfo  {journal} {Journal of Applied Physics}\ }\textbf {\bibinfo {volume} {134}} (\bibinfo {year} {2023})}\BibitemShut {NoStop}%
\bibitem [{\citenamefont {Hamamoto}\ \emph {et~al.}(2024)\citenamefont {Hamamoto}, \citenamefont {Bhunia}, \citenamefont {Bhattacharya}, \citenamefont {Takahashi},\ and\ \citenamefont {Kubo}}]{hamamoto2024dielectric}%
  \BibitemOpen
  \bibfield  {author} {\bibinfo {author} {\bibfnamefont {T.}~\bibnamefont {Hamamoto}}, \bibinfo {author} {\bibfnamefont {A.}~\bibnamefont {Bhunia}}, \bibinfo {author} {\bibfnamefont {R.~K.}\ \bibnamefont {Bhattacharya}}, \bibinfo {author} {\bibfnamefont {H.}~\bibnamefont {Takahashi}},\ and\ \bibinfo {author} {\bibfnamefont {Y.}~\bibnamefont {Kubo}},\ }\bibfield  {title} {\bibinfo {title} {Dielectric microwave resonator with large optical apertures for spin-based quantum devices},\ }\href@noop {} {\bibfield  {journal} {\bibinfo  {journal} {Applied Physics Letters}\ }\textbf {\bibinfo {volume} {124}} (\bibinfo {year} {2024})}\BibitemShut {NoStop}%
\bibitem [{\citenamefont {Fujito}(1981)}]{fujito1981magnetic}%
  \BibitemOpen
  \bibfield  {author} {\bibinfo {author} {\bibfnamefont {T.}~\bibnamefont {Fujito}},\ }\bibfield  {title} {\bibinfo {title} {Magnetic interaction in solvent-free dpph and dpph-solvent complexes.},\ }\href@noop {} {\bibfield  {journal} {\bibinfo  {journal} {Bulletin of the Chemical Society of Japan}\ }\textbf {\bibinfo {volume} {54}},\ \bibinfo {pages} {3110} (\bibinfo {year} {1981})}\BibitemShut {NoStop}%
\bibitem [{\citenamefont {Mergenthaler}\ \emph {et~al.}(2017)\citenamefont {Mergenthaler}, \citenamefont {Liu}, \citenamefont {Le~Roy}, \citenamefont {Ares}, \citenamefont {Thompson}, \citenamefont {Bogani}, \citenamefont {Luis}, \citenamefont {Blundell}, \citenamefont {Lancaster}, \citenamefont {Ardavan} \emph {et~al.}}]{mergenthaler2017strong}%
  \BibitemOpen
  \bibfield  {author} {\bibinfo {author} {\bibfnamefont {M.}~\bibnamefont {Mergenthaler}}, \bibinfo {author} {\bibfnamefont {J.}~\bibnamefont {Liu}}, \bibinfo {author} {\bibfnamefont {J.~J.}\ \bibnamefont {Le~Roy}}, \bibinfo {author} {\bibfnamefont {N.}~\bibnamefont {Ares}}, \bibinfo {author} {\bibfnamefont {A.~L.}\ \bibnamefont {Thompson}}, \bibinfo {author} {\bibfnamefont {L.}~\bibnamefont {Bogani}}, \bibinfo {author} {\bibfnamefont {F.}~\bibnamefont {Luis}}, \bibinfo {author} {\bibfnamefont {S.~J.}\ \bibnamefont {Blundell}}, \bibinfo {author} {\bibfnamefont {T.}~\bibnamefont {Lancaster}}, \bibinfo {author} {\bibfnamefont {A.}~\bibnamefont {Ardavan}}, \emph {et~al.},\ }\bibfield  {title} {\bibinfo {title} {Strong coupling of microwave photons to antiferromagnetic fluctuations in an organic magnet},\ }\href@noop {} {\bibfield  {journal} {\bibinfo  {journal} {Physical review letters}\ }\textbf {\bibinfo {volume} {119}},\ \bibinfo {pages} {147701} (\bibinfo {year} {2017})}\BibitemShut {NoStop}%
\bibitem [{\citenamefont {Diniz}\ \emph {et~al.}(2011)\citenamefont {Diniz}, \citenamefont {Portolan}, \citenamefont {Ferreira}, \citenamefont {G{\'e}rard}, \citenamefont {Bertet},\ and\ \citenamefont {Auffeves}}]{diniz2011strongly}%
  \BibitemOpen
  \bibfield  {author} {\bibinfo {author} {\bibfnamefont {I.}~\bibnamefont {Diniz}}, \bibinfo {author} {\bibfnamefont {S.}~\bibnamefont {Portolan}}, \bibinfo {author} {\bibfnamefont {R.}~\bibnamefont {Ferreira}}, \bibinfo {author} {\bibfnamefont {J.}~\bibnamefont {G{\'e}rard}}, \bibinfo {author} {\bibfnamefont {P.}~\bibnamefont {Bertet}},\ and\ \bibinfo {author} {\bibfnamefont {A.}~\bibnamefont {Auffeves}},\ }\bibfield  {title} {\bibinfo {title} {Strongly coupling a cavity to inhomogeneous ensembles of emitters: Potential for long-lived solid-state quantum memories},\ }\href@noop {} {\bibfield  {journal} {\bibinfo  {journal} {Physical Review A—Atomic, Molecular, and Optical Physics}\ }\textbf {\bibinfo {volume} {84}},\ \bibinfo {pages} {063810} (\bibinfo {year} {2011})}\BibitemShut {NoStop}%
\bibitem [{\citenamefont {Kurucz}\ \emph {et~al.}(2011)\citenamefont {Kurucz}, \citenamefont {Wesenberg},\ and\ \citenamefont {M{\o}lmer}}]{kurucz2011spectroscopic}%
  \BibitemOpen
  \bibfield  {author} {\bibinfo {author} {\bibfnamefont {Z.}~\bibnamefont {Kurucz}}, \bibinfo {author} {\bibfnamefont {J.}~\bibnamefont {Wesenberg}},\ and\ \bibinfo {author} {\bibfnamefont {K.}~\bibnamefont {M{\o}lmer}},\ }\bibfield  {title} {\bibinfo {title} {Spectroscopic properties of inhomogeneously broadened spin ensembles in a cavity},\ }\href@noop {} {\bibfield  {journal} {\bibinfo  {journal} {Physical Review A—Atomic, Molecular, and Optical Physics}\ }\textbf {\bibinfo {volume} {83}},\ \bibinfo {pages} {053852} (\bibinfo {year} {2011})}\BibitemShut {NoStop}%
\bibitem [{\citenamefont {Pechal}\ \emph {et~al.}(2014)\citenamefont {Pechal}, \citenamefont {Huthmacher}, \citenamefont {Eichler}, \citenamefont {Zeytino{\u{g}}lu}, \citenamefont {Abdumalikov~Jr}, \citenamefont {Berger}, \citenamefont {Wallraff},\ and\ \citenamefont {Filipp}}]{pechal2014microwave}%
  \BibitemOpen
  \bibfield  {author} {\bibinfo {author} {\bibfnamefont {M.}~\bibnamefont {Pechal}}, \bibinfo {author} {\bibfnamefont {L.}~\bibnamefont {Huthmacher}}, \bibinfo {author} {\bibfnamefont {C.}~\bibnamefont {Eichler}}, \bibinfo {author} {\bibfnamefont {S.}~\bibnamefont {Zeytino{\u{g}}lu}}, \bibinfo {author} {\bibfnamefont {A.~A.}\ \bibnamefont {Abdumalikov~Jr}}, \bibinfo {author} {\bibfnamefont {S.}~\bibnamefont {Berger}}, \bibinfo {author} {\bibfnamefont {A.}~\bibnamefont {Wallraff}},\ and\ \bibinfo {author} {\bibfnamefont {S.}~\bibnamefont {Filipp}},\ }\bibfield  {title} {\bibinfo {title} {Microwave-controlled generation of shaped single photons in circuit quantum electrodynamics},\ }\href@noop {} {\bibfield  {journal} {\bibinfo  {journal} {Physical Review X}\ }\textbf {\bibinfo {volume} {4}},\ \bibinfo {pages} {041010} (\bibinfo {year} {2014})}\BibitemShut {NoStop}%
\bibitem [{\citenamefont {Couillard}(2020)}]{couillard_keysight_E5071C}%
  \BibitemOpen
  \bibfield  {author} {\bibinfo {author} {\bibfnamefont {M.}~\bibnamefont {Couillard}},\ }\href@noop {} {\bibinfo {title} {keysight\_{E}5071{C}}},\ \bibinfo {howpublished} {\url{https://github.com/mathieu-couillard/keysight_E5071C}} (\bibinfo {year} {2020}),\ \bibinfo {note} {version 1.0.0. Python driver for the Keysight E5071C Vector Network Analyzer (VNA).}\BibitemShut {Stop}%
\bibitem [{\citenamefont {Couillard}(2025)}]{couillard_hqd_epro}%
  \BibitemOpen
  \bibfield  {author} {\bibinfo {author} {\bibfnamefont {M.}~\bibnamefont {Couillard}},\ }\href@noop {} {\bibinfo {title} {hqd\_epr}},\ \bibinfo {howpublished} {\url{https://github.com/mathieu-couillard/hqd_epr}} (\bibinfo {year} {2025}),\ \bibinfo {note} {version 1.0.0. Private GitHub repository. Scripts to control and analyse data from a VNA and OPX+ for cryogenic EPR.}\BibitemShut {Stop}%
\bibitem [{\citenamefont {{Supercon Inc.}}(2026)}]{Supercon54S43Spec}%
  \BibitemOpen
  \bibfield  {author} {\bibinfo {author} {\bibnamefont {{Supercon Inc.}}},\ }\href {http://www.supercon-wire.com/content/nbti-superconducting-wires} {\emph {\bibinfo {title} {Multifilament {NbTi} Superconducting Wire: {54S43} Technical Specifications}}},\ \bibinfo {organization} {Supercon Inc.},\ \bibinfo {address} {Shrewsbury, MA} (\bibinfo {year} {2026}),\ \bibinfo {note} {standard 54-filament NbTi conductor, Cu:SC ratio 1.3:1}\BibitemShut {NoStop}%
\bibitem [{\citenamefont {Grobet}\ \emph {et~al.}(1978)\citenamefont {Grobet}, \citenamefont {Van~Gerven},\ and\ \citenamefont {Van~den Bosch}}]{Grobet1978spin}%
  \BibitemOpen
  \bibfield  {author} {\bibinfo {author} {\bibfnamefont {P.}~\bibnamefont {Grobet}}, \bibinfo {author} {\bibfnamefont {L.}~\bibnamefont {Van~Gerven}},\ and\ \bibinfo {author} {\bibfnamefont {A.}~\bibnamefont {Van~den Bosch}},\ }\bibfield  {title} {\bibinfo {title} {The spin magnetism of $\alpha$, '‐diphenyl‐$\beta$‐picrylhydrazyl (dpph)},\ }\href {https://doi.org/10.1063/1.435589} {\bibfield  {journal} {\bibinfo  {journal} {The Journal of Chemical Physics}\ }\textbf {\bibinfo {volume} {68}},\ \bibinfo {pages} {5225} (\bibinfo {year} {1978})}\BibitemShut {NoStop}%
\bibitem [{\citenamefont {Prokhorov}\ and\ \citenamefont {Fedorov}(1963)}]{prokhorov1963antiferromagnetism}%
  \BibitemOpen
  \bibfield  {author} {\bibinfo {author} {\bibfnamefont {A.}~\bibnamefont {Prokhorov}}\ and\ \bibinfo {author} {\bibfnamefont {V.}~\bibnamefont {Fedorov}},\ }\bibfield  {title} {\bibinfo {title} {Antiferromagnetism of free radicals},\ }\href@noop {} {\bibfield  {journal} {\bibinfo  {journal} {Soviet Physics JETP}\ }\textbf {\bibinfo {volume} {16}} (\bibinfo {year} {1963})}\BibitemShut {NoStop}%
\bibitem [{\citenamefont {Bleaney}\ and\ \citenamefont {Bowers}(1952)}]{bleaney1952anomalous}%
  \BibitemOpen
  \bibfield  {author} {\bibinfo {author} {\bibfnamefont {B.}~\bibnamefont {Bleaney}}\ and\ \bibinfo {author} {\bibfnamefont {K.}~\bibnamefont {Bowers}},\ }\bibfield  {title} {\bibinfo {title} {Anomalous paramagnetism of copper acetate},\ }\href@noop {} {\bibfield  {journal} {\bibinfo  {journal} {Proceedings of the Royal Society of London. Series A. Mathematical and Physical Sciences}\ }\textbf {\bibinfo {volume} {214}},\ \bibinfo {pages} {451} (\bibinfo {year} {1952})}\BibitemShut {NoStop}%
\bibitem [{\citenamefont {Reif}(2009)}]{reif2009fundamentals}%
  \BibitemOpen
  \bibfield  {author} {\bibinfo {author} {\bibfnamefont {F.}~\bibnamefont {Reif}},\ }\href@noop {} {\emph {\bibinfo {title} {Fundamentals of statistical and thermal physics}}}\ (\bibinfo  {publisher} {Waveland Press},\ \bibinfo {year} {2009})\BibitemShut {NoStop}%
\bibitem [{\citenamefont {Couillard}(2024)}]{couillardFidSimulation}%
  \BibitemOpen
  \bibfield  {author} {\bibinfo {author} {\bibfnamefont {M.}~\bibnamefont {Couillard}},\ }\href@noop {} {\bibinfo {title} {fid\_simulation}},\ \bibinfo {howpublished} {\url{https://github.com/mathieu-couillard/fid_simulation}} (\bibinfo {year} {2024})\BibitemShut {NoStop}%
\end{thebibliography}%
\end{document}